\documentclass[]{aastex63}

\newcommand{\sdo}{\textit{SDO}}
\newcommand{\aia}{\textit{SDO}/AIA}

\newcommand{\hinode}{\textit{Hinode}}

\newcommand{\magixs}{\textit{MaGIXS}}

\newcommand{\nustar}{\textit{NuSTAR}}

\newcommand{\iris}{\textit{IRIS}}

\usepackage{graphicx}
\usepackage{amssymb}                    
\usepackage{ulem}
\usepackage{color}
\usepackage{wrapfig}
\usepackage{sidecap}
\usepackage{booktabs}
\usepackage{longtable}
\usepackage{natbib}
\usepackage{hyperref}
\usepackage{enumerate}
\usepackage{appendix}

\graphicspath{figures/}

\shorttitle{MaGIXS Mission Science}
\shortauthors{Savage et. al}
 
\begin{document}

\title{The First Flight of the Marshall Grazing Incidence X-ray Spectrometer (MaGIXS)}

\newcommand{\msfc}{NASA Marshall Space Flight Center, Huntsville, AL 35812, USA}
\newcommand{\uah}{Center for Space Plasma and Aeronomic Research, University of Alabama Huntsville, Huntsville, AL, 35812, USA}
\newcommand{\izentis}{Izentis LLC, Cambridge, MA 02139, USA}
\newcommand{\sao}{Center for Astrophysics $\vert$\ Harvard\ \&\ Smithsonian, Cambridge, MA 02138, USA}
\newcommand{\themit}{Massachusetts Institute of Technology, Cambridge, MA 02139, USA}
\newcommand{\rice}{Department of Physics and Astronomy (MS 108), Rice University, 6100 Main Street, Houston, TX 77005, USA}
\newcommand{\uclan}{University of Central Lancashire, Preston, PR1 2HE, UK}
\newcommand{\ucamb}{DAMTP, Centre for Mathematical Sciences, Wilberforce Road, Cambridge CB3, UK}
\newcommand{\nrl}{Space Science Division, Naval Research Laboratory, Washington, DC 20375, USA}
\newcommand{\wedden}{Weddendorf Design, Inc., Huntsville, AL 35803, USA}
\newcommand{\bwxt}{BWX Technologies, Lynchburg, VA 24504, USA}

\correspondingauthor{Sabrina Savage}
\email{sabrina.savage@nasa.gov}

\author[0000-0002-6172-0517]{Sabrina L. Savage}\affiliation{\msfc}
\author[0000-0002-5608-531X]{Amy R. Winebarger}\affiliation{\msfc}
\author[0000-0003-1057-7113]{Ken Kobayashi}\affiliation{\msfc}
\author[0000-0002-4454-147X]{P.S. Athiray}\affiliation{\uah}\affiliation{\msfc}
\author{Dyana Beabout}\affiliation{\msfc}
\author[0000-0001-9638-3082]{Leon Golub}\affiliation{\sao}
\author[0000-0002-1025-9863]{Robert W. Walsh}\affiliation{\uclan}
\author{Brent Beabout}\affiliation{\msfc}
\author[0000-0002-3300-6041]{Stephen Bradshaw}\affiliation{\rice}
\author[0000-0001-5927-3300]{Alexander R. Bruccoleri}\affiliation{\izentis}
\author[0000-0002-7139-6191]{Patrick R. Champey}\affiliation{\msfc}
\author{Peter Cheimets}\affiliation{\sao}
\author{Jonathan Cirtain}\affiliation{\bwxt}
\author[0000-0001-7416-2895]{Edward E. DeLuca}\affiliation{\sao}
\author[0000-0002-4125-0204]{Giulio Del Zanna}\affiliation{\ucamb}
\author{Anthony Guillory}\affiliation{\msfc}
\author{Harlan Haight}\affiliation{\msfc}
\author[0000-0001-9980-5295]{Ralf K.\ Heilmann}\affiliation{\themit}
\author[0000-0002-6747-9648]{Edward Hertz}\affiliation{\sao}
\author{William Hogue}\affiliation{\msfc}
\author{Jeffery Kegley}\affiliation{\msfc}
\author{Jeffery Kolodziejczak}\affiliation{\msfc}
\author[0000-0001-8775-913X]{Chad Madsen}\affiliation{\sao}
\author[0000-0002-6418-7914]{Helen Mason}\affiliation{\ucamb}
\author[0000-0002-9921-7757]{David E. McKenzie}\affiliation{\msfc}
\author{Jagan Ranganathan}\affiliation{\msfc}
\author[0000-0002-6903-6832]{Katharine K. Reeves}\affiliation{\sao}
\author{Bryan Robertson}\affiliation{\msfc}
\author[0000-0001-6932-2612]{Mark L. Schattenburg}\affiliation{\themit}
\author{Jorg Scholvin}\affiliation{\izentis}
\author{Richard Siler}\affiliation{\msfc}
\author[0000-0002-0405-0668]{Paola Testa}\affiliation{\sao}
\author[0000-0002-7219-1526]{Genevieve D.\ Vigil}\affiliation{\msfc}
\author[0000-0001-6102-6851]{Harry P. Warren}\affiliation{\nrl}
\author{Benjamin Watkinson}\affiliation{\uclan}
\author{Bruce Weddendorf}\affiliation{\wedden}
\author{Ernest Wright}\affiliation{\msfc}

\begin{abstract}
    
The Marshall Grazing Incidence X-ray Spectrometer (\magixs) sounding rocket experiment launched on July 30, 2021 from the White Sands Missile Range in New Mexico. \magixs\ is a unique solar observing telescope developed to capture X-ray spectral images, in the 6\,--\,24\,\AA\ wavelength range, of coronal active regions.  Its novel design takes advantage of recent technological advances related to fabricating and optimizing X-ray optical systems as well as breakthroughs in inversion methodologies necessary to create spectrally pure maps from overlapping spectral images.  \magixs\ is the first instrument of its kind to provide spatially resolved soft X-ray spectra across a wide field of view. The plasma diagnostics available in this spectral regime make this instrument a powerful tool for probing solar coronal heating.  This paper presents details from the first \magixs\ flight, the captured observations, the data processing and inversion techniques, and the first science results.

\end{abstract}
\section{\label{introduction}Introduction}

\begin{figure}[!ht]
\centering\includegraphics[width=.75\textwidth]{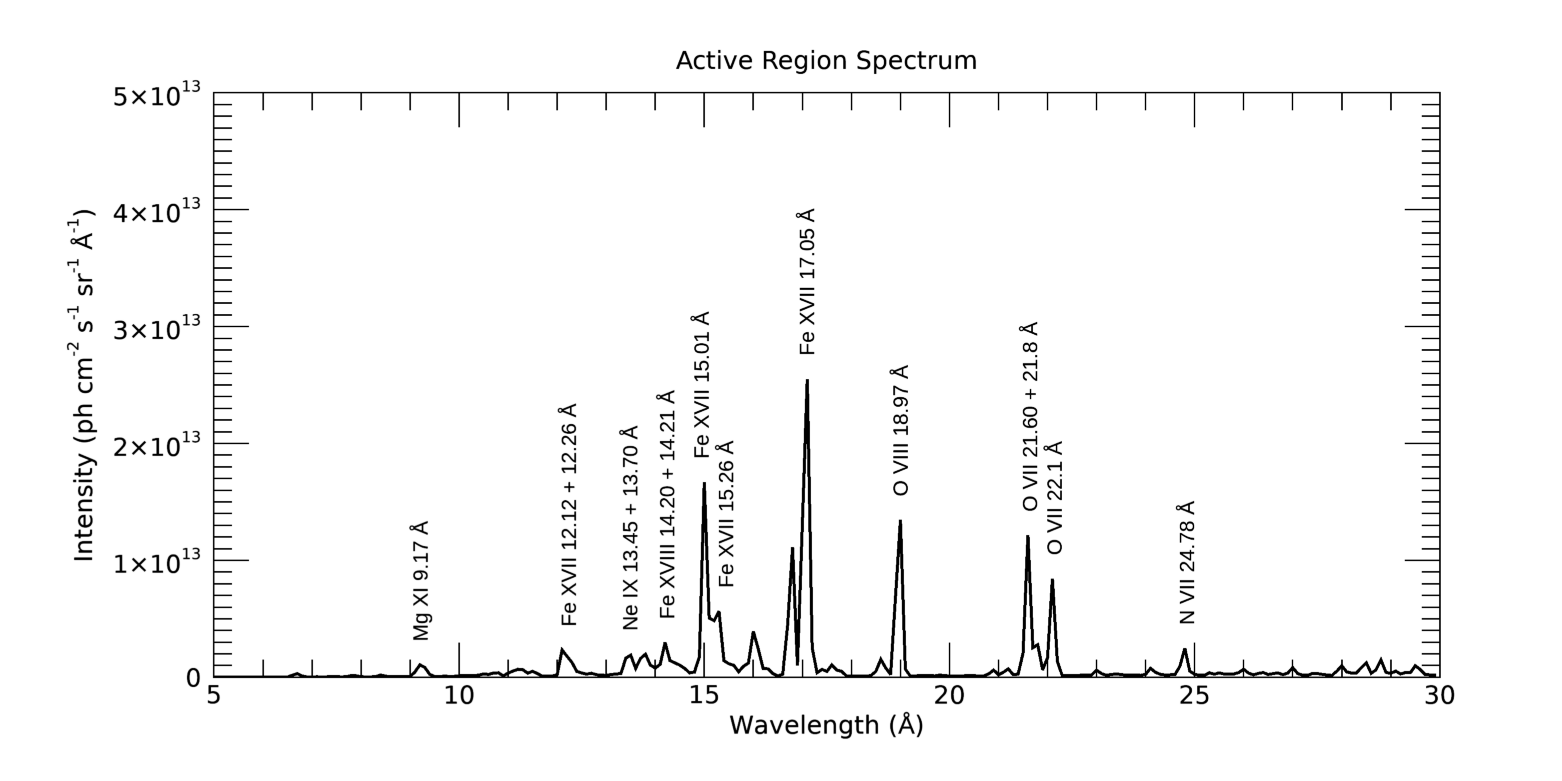}
\caption{The \magixs\ wavelength range (6\,--\,24\,\AA) and spectrum with the strongest spectral lines identified, derived from CHIANTI using an active region differential emission measure.}
\label{fig:wavelength_range}
\end{figure}

X-ray spectroscopy provides unique capabilities for answering fundamental questions in solar physics \citep{2021FrASS...8...33D, 2021FrASS...8...50Y}.  The X-ray regime is dominated by emission lines formed at high temperatures, with untapped potential to yield insights into basic physical processes of the Sun and stars that are not accessible by any other means.  The Marshall Grazing Incidence X-ray Spectrometer (\magixs) is a sounding rocket instrument developed as a pathfinder to acquire the first ever spatially and spectrally resolved images discriminating between coronal active region structures in X-rays, without the restriction of a slit. The \magixs\ wavelength range (6\,--\,24\,\AA) with available spectral lines is shown in Figure~\ref{fig:wavelength_range}.  Herein, we refer to these wavelengths as soft X-rays (SXRs), as compared to hard X-ray detectors such as the Reuven Ramaty High Energy Solar Spectroscopic Imager (RHESSI); note, however, that this classification varies \cite[e.g.,][]{2021FrASS...8...33D}.

For the last 20 years, solar astrophysics has heavily relied on measurements of coronal plasma using extreme ultraviolet (EUV), ultraviolet (UV), or white light instrumentation along with broadband X-ray imaging. These resources provide limited spectral information for measuring the temperature, density, or element fractionation for the various structures and events in the corona. As there has been hardly any SXR imaging spectroscopy of the solar corona since the {\it Orbiting Solar Observatory} (OSO) era (1962--1975), besides the Bragg Flat Crystal Spectrometer (FCS) instrument on the {\it Solar Maximum Mission} (SMM, 1980--1989), which had relatively poor spatial resolution, there is a massive well of untapped potential for future discovery.  The first flight of \magixs, occurring after a decade of technology and instrument development, demonstrates a revolutionary concept for advancing such grazing incidence imaging spectroscopy.

For a short-duration rocket flight, the overarching science goal of \magixs\ targets the frequency of heating in active region cores, an essential measurement needed to solve the elusive coronal heating problem discovered by Edl\'{e}n and Grotrian in the 1930's.  Two primary mechanisms are anticipated to play dominant roles in the transfer and dissipation of energy into the corona -- magnetic reconnection \citep{1983ApJ...264..635P,1983ApJ...264..642P} and wave heating \citep{2011ApJ...736....3V,2014ApJ...787...87V}.  Energy released through single field line reconnection events (i.e., nanoflares) due to forced interactions between stressed magnetic fields via photospheric motions is expected to be sporadic, short-lived, and \textit{infrequent} \citep[e.g,][]{2010ApJ...719..591L}.  Conversely, magnetic wave heating along these field lines would be sporadic, short-lived, but \textit{frequent} \citep{2012ApJ...746...81A}.  

\begin{figure}[t!]
\centering\includegraphics[width=.6\textwidth]{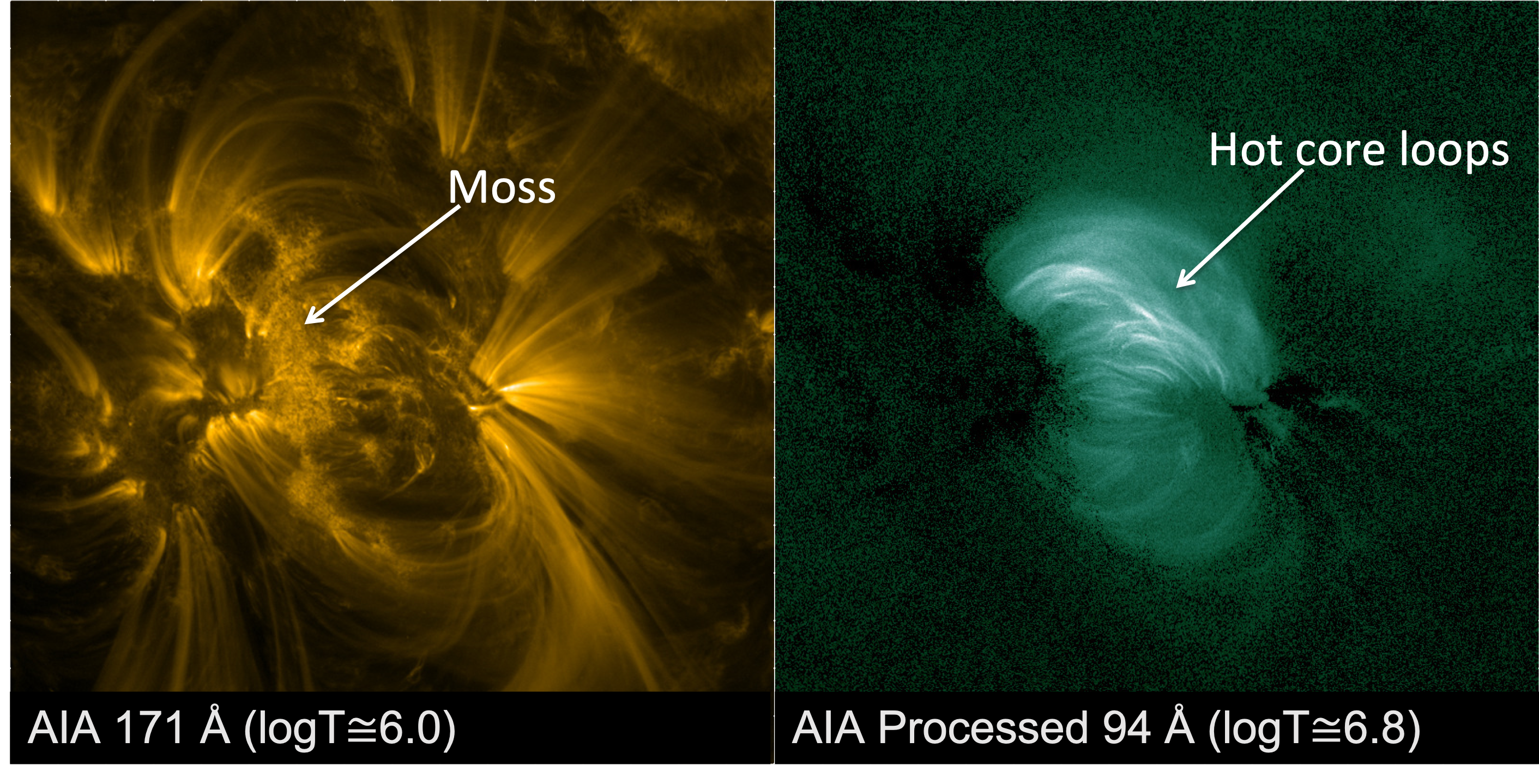}
\caption{(Left) Active Region 11339 observed Nov 10, 2011.  The footpoints of the high temperature loops form the reticulated emission in the \sdo/AIA Fe~IX/X 171\,\AA\ image.  (Right) The hot active region core is shown in the processed {\it SDO}/AIA Fe~XVIII 94\,\AA\ image.  The size of the image is 8\arcmin$\times$8\arcmin.  The log of the approximate dominant emitting temperature is given in each image.}
\label{fig:aia_111110}
\end{figure}

The heating frequency in the highest temperature loops in the solar corona, those in the active region core, remains the most controversial.  Figure~\ref{fig:aia_111110} shows an example active region observed with the {\it Solar Dynamics Observatory} (SDO) Atmospheric Imaging Assembly (AIA). The left panel shows the 1\,MK footpoints of high-temperature loops, or ``moss,'' and the right panel shows the 94\,\AA\ channel emission with the cool contribution removed \citep[following][]{2012ApJ...759..141W}.  The remaining emission is expected to be from the Fe~XVIII spectral line formed within the $\sim$\,4--8\,MK range \citep{Testa2020,Testa2012,Reale2019}.   Many diagnostics have attempted to glean the heating frequency from cooler (1\,--\,4\,MK) observations of these hot core loops \citep[see, for instance,][]{2014ApJ...784...49C,2016ApJ...821...63B,2019ApJ...880...56B,2021ApJ...919..132B}; however, these are often difficult to interpret due to contributions of overlying cool structures or the moss footpoints \citep[e.g.,][]{2010ApJ...723..713T} and other errors \citep{2013ApJ...774...31G}. 

Spectral observations in the SXR regime provide the most unambiguous means of differentiating between the heating frequency cases through characteristic variations in observables. \magixs\ is designed to optimally discriminate between these competing coronal heating theories by providing SXR spectra along spatially-resolved active region features -- \textbf{a combination that no other solar instrument offers}. The unique and powerful plasma diagnostics afforded through SXR spectral imaging to achieve this goal include measurements of:


\begin{itemize}
    \item \vspace{-8pt}emission from Fe~XVII\,--\,Fe~XX to assess the temperature distribution above 4\,MK in an active region,
    \item \vspace{-8pt}elemental abundances of high temperature plasma,
    \item \vspace{-8pt}the temporal variability of high temperature plasma (Fe~XVII) at high cadence, 
    \item \vspace{-8pt}the presence of non-Maxwellian electron distributions.
\end{itemize}

\noindent \textbf{I. High-temperature, low-emission plasma:} \textit{High-frequency heating scenario:} If the frequency of energy release on a given strand\footnote{Here we use the term ``strand'' to refer to the fundamental plasma feature in the corona and the term ``loop'' to refer to a spatially coherent structure in an observation.  A loop can consist of a single strand, or, more likely,  many, sub-resolution strands.} is high, the plasma along that strand does not have time to cool before being re-heated.  As a result, the temperature and density of the strand remain relatively constant.  \textit{Low-frequency heating scenario:} If the frequency of heating events is low (i.e., the time between two heating events on a given strand is longer than the plasma's cooling time), the plasma's density and temperature along that strand would be dynamic and evolving.  During its evolution, the temperature would be both much higher and much lower than the average temperature.  Because an observed loop is almost certainly formed of many strands \citep[e.g.,][]{2014ApJ...786...82K,2020ApJ...902...90W}, the loop's properties may or may not reflect this evolution.   If the loop is formed of many unresolved strands, each strand being heated randomly and then evolving, the observed loop's intensity can appear steady regardless of the dynamic nature of the plasma along a single strand \citep{2009ASPC..415..221K}.   

One well-studied observation that can discriminate between low- and high-frequency heating in active region cores is the relative amount of high-temperature ($\sim$\,5--10\,MK) to average temperature ($\sim$\,3--5\,MK) plasma \citep[e.g.,][]{Reale2009,McTiernan2009,Ko2009,Sylwester2010,Testa2011,Miceli2012,Petralia14,Brosius2014,Ishikawa2014,Athiray2020b,2016ApJ...829...31B}.  \cite{athiray2019} demonstrated that the heating frequency can be easily gleaned from simple intensity ratios between spectrally pure Fe~XVII, XVIII, or XIX intensities (Figure 11 therein).  Further, they demonstrated that intensities from existing instruments, like \emph{Hinode's} X-Ray Telescope (XRT) or AIA, were \emph{insensitive} to the heating frequency, possibly due to their broadband filter response functions. The relative variations in the responses may not have the resolution to distinguish high temperature emission.  Instead, these instruments are more sensitive to the cool slope ($\alpha$) versus the hot slope ($\beta$). 

\begin{figure}[!ht]
\centering\includegraphics[width=.5\textwidth]{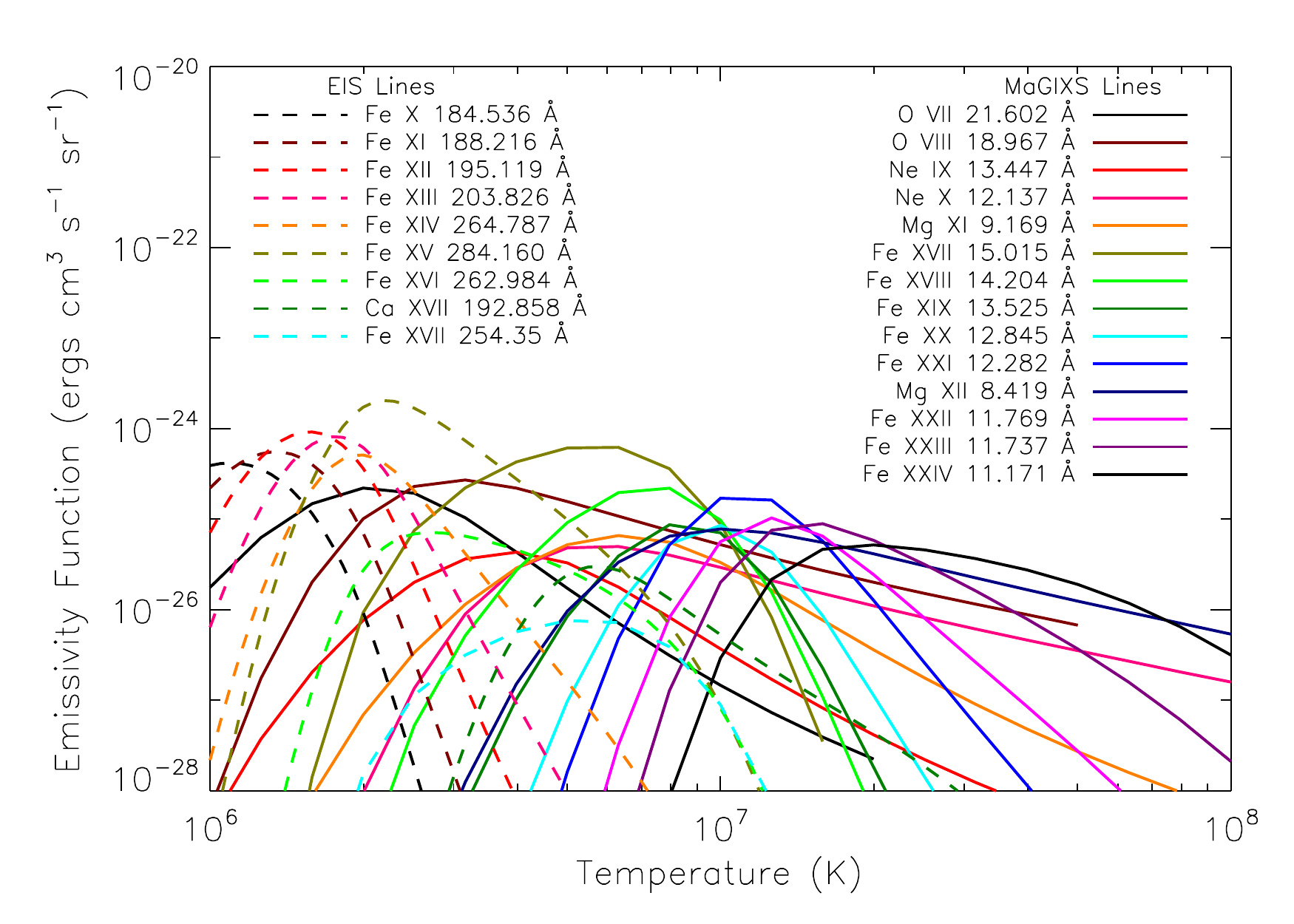}
\caption{The emissivity functions for key MaGIXS spectral lines (solid) compared to the emissivity functions of key \hinode/EIS spectral lines (dashed). Though the EIS wavelength range does contain a few high temperature lines, these are weak and blended and tend only to be well observed during a solar flare.  The temperature range of $6.2\,<$\,Log\,T$\,< 7.2$ is well covered by the \magixs\ spectral lines.}
\label{fig:emis_func_euv}
\end{figure}

The emissivity functions for key \magixs\ strong spectral lines are shown in Figure~1 of \cite{champey2022}.  \magixs\ provides better high temperature coverage and temperature discrimination than is currently available in EUV spectrometers or in EUV or X-ray imagers (see Figure~\ref{fig:emis_func_euv} for comparative examples to \hinode's EUV Imaging Spectrograph (EIS)).  The EIS instrument, which currently provides the highest temperature spectrally pure measurements of the solar corona, is able to detect plasma with temperatures up to 4\,MK very well, but in the 4--10 MK range the lines are weak and blended with other transitions \citep[see, e.g.][]{delzanna_etal_flare:11,delzanna_ishikawa:09}. \\

\noindent \textbf{II. Element abundances of high temperature plasma:}  Early spectroscopic observations at X-ray and EUV wavelengths allowed the first measurements of the abundances of different elements in the solar corona.  The composition of the corona, however, unexpectedly did not always match the composition of the underlying photosphere (e.g., \cite{Evans68,Widing68,Withbroe75,Parkinson77,Veck81}). The abundances of a few elements sometimes appeared to be enhanced, while the abundances of other elements remained closer to their photospheric values, or even lower \citep{raymond_etal:97}.  The enhanced elements, such as Fe and Si, have low first ionization potential (FIP $<\,10$~eV), while the non-enhanced elements, such as O and Ne, have high-FIP [$\gtrsim$\,10\,eV; see reviews by \cite{Meyer85,Feldman92,Sylwester10,Testa2010,laming:2015,delzanna_mason:2018}].  The ``FIP bias'' (i.e., the enhancement ratio of low FIP elements) is generally found to be a factor of $\sim$2\,--\,4.


Several studies have revealed that FIP bias of coronal structures is a consequence of the plasma's time of confinement in coronal structures (e.g., \cite{warren2014,delzanna2014,ugarte2012}.  Impulsively heated loops or jets have a photospheric composition (low FIP bias), while quiescent loops have coronal abundances (high FIP bias), thereby linking abundance measurements to the frequency of coronal heating. The \magixs\ spectral region is the most suitable to measure the FIP bias of hot plasma in the 3--10 MK range. Previous observations in this spectral region of quiescent active region cores have indicated a FIP bias of about 3 \citep{delzanna2014}.  Simple ratios of the the strong lines listed in Table~\ref{tab:key_abund_lines} provide the relative abundance diagnostics largely independent 
of the temperature structure of the plasma (e.g., \citealt{DrakeTesta05,Huenemoerder09,Testa2010}).  These ratios allow for comparisons to expected abundance measurements from modeled heating frequency scenarios and, by virtue of \magixs' spatial resolution, comparisons can be made between active region structures.  \\

\begin{deluxetable}{lcc}
\tabletypesize{\small}
\tablewidth{6in}
\tablecolumns{3}
\tablecaption{Key spectral lines appropriate for abundance analysis. \label{tab:key_abund_lines}}
\tablehead{\colhead{Ion} & \colhead{Wavelength} & \colhead{Log Temperature}\\
\colhead{} & \colhead{[\AA]} & \colhead{}}
\startdata
Mg \sc{XII} & 8.42\,\AA & 6.9\\
Mg \sc{XI} & 9.16\,\AA & 6.4\\
Ne \sc{X} & 12.13\,\AA & 6.6 \\
Ne \sc{IX} & 13.45\,\AA & 6.2\\
Fe \sc{XVIII} & 14.21\,\AA & 6.8 \\
Fe \sc{XVII} & 15.01\,\AA & 6.6 \\
O \sc{VIII} & 18.97\,\AA & 6.4 \\
O \sc{VII} & 21.60\,\AA & 6.3 \\
\enddata
\end{deluxetable}

\noindent \textbf{III. Temporal variability of high temperature plasma:} In an active region, the overlapping of many optically thin loops, possibly consisting of smaller unresolved strands \citep[e.g.][]{Brooks2012}, complicates the identification of individual heating events.  Nevertheless, statistical analysis of high temperature light curves can provide information on the frequency of heating in an active region \citep[e.g.,][]{ugarte2014}.  Since the light curve of an individual impulsive heating event includes a steep rise in time followed by a slower decay phase \citep{lopez2010}, an impulsive heating scenario would result in significant skew in the light curve fluctuations as a function of time.  An analysis performed by \citet{Terzo2011} found a skewness in XRT active region light curves that could not be accounted for by Poisson noise alone.  The lifetime of their identified events were on the order of 100-500\,s, and the XRT intensity enhancement was on the order of 20\%.  However, these measurements are very sensitive to the noise in the data, and the light curve identification is further complicated by the broad temperature response of the XRT filters.  Additionally, the measurements by \cite{Terzo2011} were performed with the XRT Al-poly filter, which has significant contributions from cooler temperatures \citep{Golub2007}. \\

\noindent \textbf{IV. Presence of non-Maxwellian electron distributions:} Departures from a Maxwellian distribution, especially the presence of high-energy tails that would be detectable in the SXR regime, are expected to arise due to magnetic reconnection or wave-particle interactions. Quantifying the number of high-energy particles provides strong constraints on the possible coronal heating mechanisms with the presence of non-Maxwellian distributions clearly indicating nanoflare heating.  The potential of the \magixs\ instrument in this respect is discussed in \cite{Jaro2019}. \\

Despite its clear utility, obtaining SXR spectra is considerably more difficult than for longer wavelengths due to the challenges involved with 1) aligning grazing incidence optics with a slit and grating assembly, 2) low SXR throughput through a slit system, and 3) fabricating a grazing incidence, varied-line space grating.  Alignment and throughput requirements are loosened (although not eliminated) with the use of a wide slot versus a slit, but with the added complication of obtaining overlapping spectral line images of the field of view across the detector.  These resulting spectroheliogram images (also referred to as ``overlappograms"), provide a wealth of information with spectra obtained across the entire field of view in a single image, but require significant advancements in deconvolution techniques in order to separate the lines.  

Fortunately, great strides have been made in recent years in advancing the necessary inversion methods for other missions (e.g., the Multi-slit Solar Explorer (MUSE)) that now make it possible to extract pure spectral line images from the \magixs\ spectroheliograms \citep{2015ApJ...807..143C,2019ApJ...882...13C,2019ApJ...882...12W}. \textbf{\magixs\ takes advantage of these revolutionary analysis breakthroughs, along with innovative advancements in high-resolution grazing incidence mirror fabrication, optimized grating lithography, and improved camera efficiencies, to produce wide field of view SXR spectral images.}

This paper provides an overview of the first \magixs\ flight, the calibration and inversion processing of the flight data, and initial results from analysis on a bright region observed by \magixs.


\section{\label{instrument}Instrument Overview}

\begin{figure}[!h]
    \centering
    \includegraphics[width=0.8\textwidth]{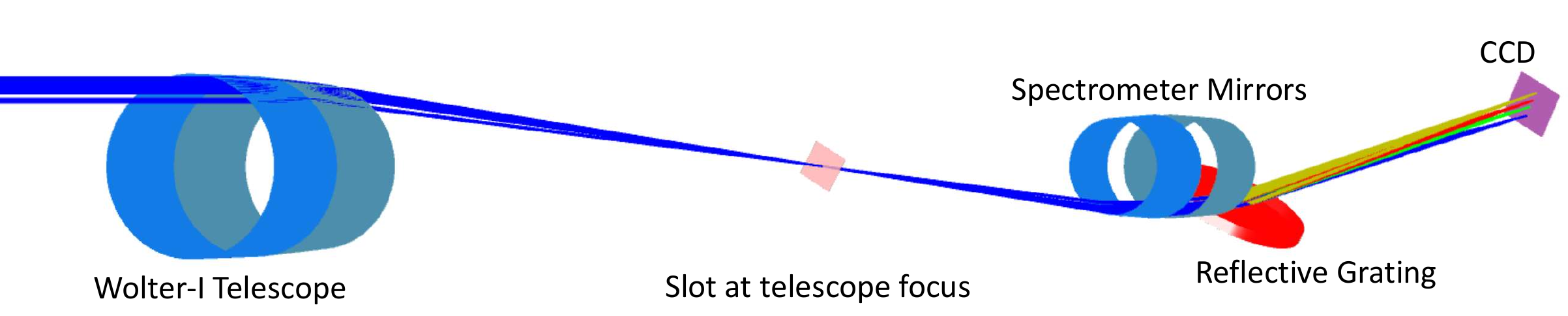}
    \caption{Ray trace diagram of the MaGIXS optical system.}
    \label{fig:raytrace}
\end{figure}

The MaGIXS instrument is described in detail in \cite{champey2022}. 
It was designed as a fully grazing-incidence slit spectrograph, consisting of a Wolter-I telescope, slit, spectrometer, CCD camera, and slit-jaw context imager. The optical path is illustrated in Figure~\ref{fig:raytrace}.
The spectrometer comprises a matched pair of grazing-incidence parabolic mirrors which re-images the slit, and a planar varied-line space grating. 
All mirrors are single, thin-shell nickel–cobalt replicated mirrors made by NASA Marshall Space Flight Center (MSFC).  The X-ray mirrors and grating are mounted on an optical bench, and are collectively termed as the Telescope Mirror Assembly (TMA) and Spectrometer Optical Assembly (SOA) (see \cite{champey2022}).
The 2k$\times$2k CCD is operated as a 2k$\times$1k frame-transfer device, allowing image readout concurrent with exposure. 
Due to the lower-than-expected throughput of the optics, as well as the recognition of 
the value of slot spectrographs in providing both spatial and spectral information, MaGIXS
was fitted with a 12'-wide slot instead of the originally intended narrow slit. 
\magixs\ includes a ``slit jaw" context imager, described in \cite{vigil2021}.  

\section{Flight overview}

\begin{figure}[!h]
    \centering
    \includegraphics[width=1\textwidth]{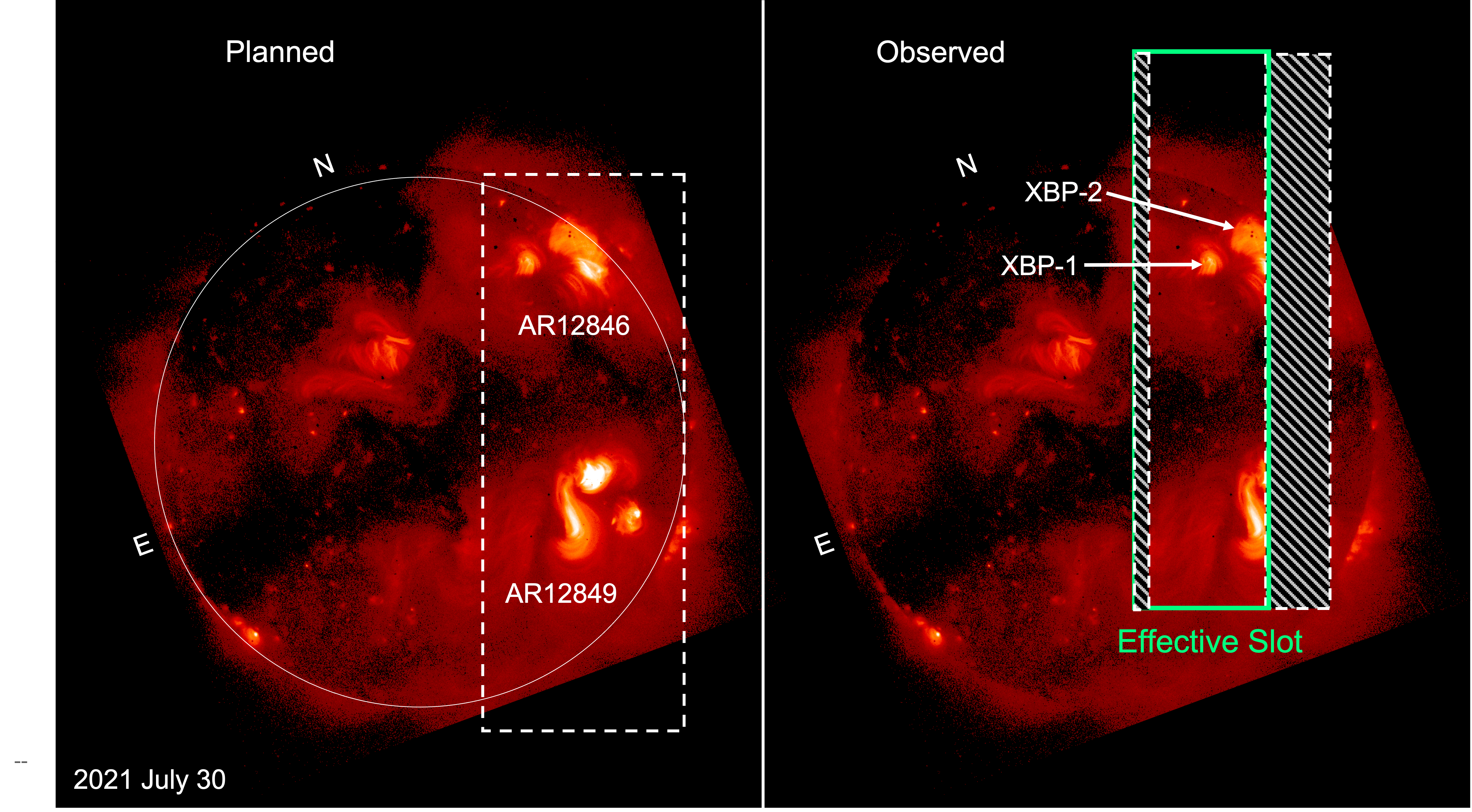}
    \caption{(Left) Planned targeted field of view for flight overlaid on a SXR image from \hinode/XRT (Thin-Be). The \magixs\ slot is indicated by the dashed box. (Right) Approximate final slot position during flight. The effective slot is a consequence of slot misalignment and vignetting (described in Section~\ref{subsec:rollpointing}). Initial analyses performed on the two labeled X-ray bright points are described in Section~\ref{discussion}.}
    \label{fig:target}
\end{figure}

The first flight of \magixs\  occurred at 18:20~UT on 2021 July 30 from the White Sands Missile Range.  With its large 12'x33' slot, \magixs\ targeted two active regions: 12846 and 12849 (Figure~\ref{fig:target} (left)). During flight, the slitjaw context imager was deemed too saturated from scattered light to be used for visual acquisition of the targets, so predetermined coordinates were relied upon.  After initial pointing adjustments were made to confirm the lack of real-time targeting capability with the slitjaw, the Solar Pointing Attitude Rocket Control System (SPARCS) maintained a constant target for the duration of the flight.  \magixs\ captured 374 total seconds of stable solar viewing data, including the intial repointing period. The final target was observed for 298 seconds.  Post-flight analysis revealed an offset between the slot and the optical axis, resulting in signficant vignetting of the system.  This effect is discussed in detail in Section~\ref{subsec:rollpointing}.  The resulting effective field of view is shown in Figure~\ref{fig:target} (right).  The X-ray bright points used for further analysis in the following sections are also highlighted. 

\begin{figure}[!ht]
    \centering
    \includegraphics[width=0.4\textwidth]{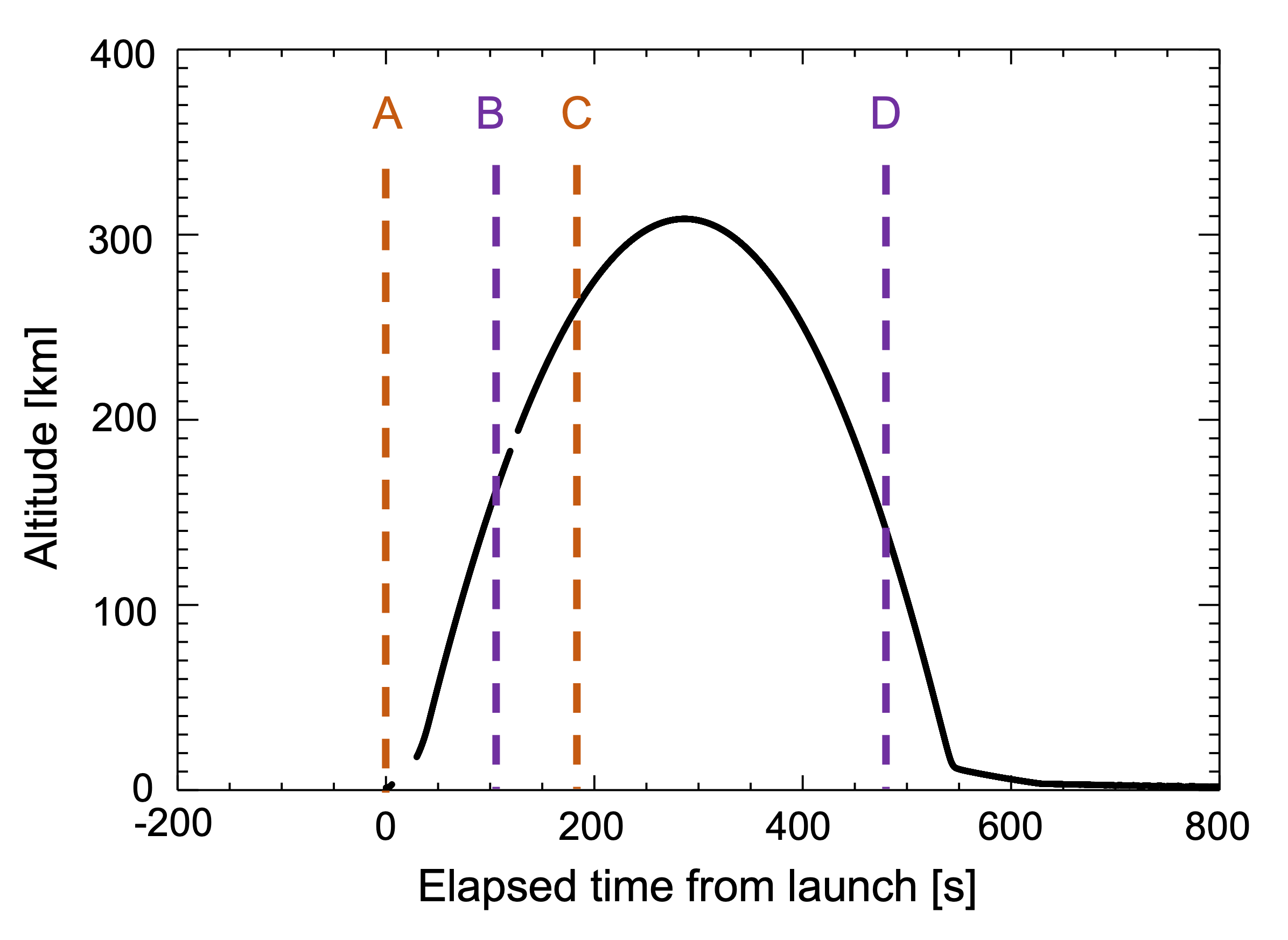}
    \caption{The altitude of the \magixs\ rocket determined from White Sands Missile Range radar data as a function of elapsed time from launch (18:20:00 UT) on 2021 July 30. [A: Launch; B: Stable pointing acquired / First SPARCS pointing; C: Final pointing maneuver; D: Stable pointing ended]}
    \label{fig:flight_profile}
\end{figure}

The altitude of the sounding rocket as a function of time as determined from White Sands Missile Range radar measurements is shown in Figure~\ref{fig:flight_profile}, with approximate flight event timings overlaid.  Table~\ref{tab:pointing} lists the times and positions of the repointings.

\begin{deluxetable}{cccc}
\tabletypesize{\small}
\tablewidth{6in}
\tablecolumns{4}
\tablecaption{SPARCS pointing coordinates from Sun center. \label{tab:pointing}}
\tablehead{\colhead{Pointing}&\colhead{Time lapsed since launch}& \colhead{North} & \colhead{West} \\
\colhead{number} & \colhead{sec}& \colhead{arcsec} & \colhead{arcsec}}
\startdata
1&T0+109 & 527.57 & -213.15\\
2&T0+133 & 332.0 & 247.68\\
3&T0+144 & 386.63 & 270.89\\
4&T0+149 & 441.30 & 294.10 \\
5&T0+155 & 495.98 & 317.31 \\
6&T0+160 & 550.65 & 340.52 \\
7&T0+182 & 605.33 & 363.72 \\
8&T0+185 & 660.00 &386.93 \\
\enddata
\end{deluxetable}

\pagebreak
\section{\textit{MaGIXS} Data Analysis}

\subsection{Data Description}

All 16-bit images acquired from the \magixs\ science camera were saved onboard in the form of FITS files.
Images were acquired at a constant 2-second cadence, starting at launch and ending after the shutter door was closed; the images acquired with the door closed are used as dark frames. Due to the utilization of frame transfer mode, there was no time gap between images.
A total of 254 frames were captured during flight, including 23 dark frames before shutter door opened, 39 frames during pointing maneuvers, 148 frames at the final stable pointing, and 10 dark frames after the shutter door closed. Each readout register of the detector included 50 Non-Active Pixels (NAPs), which are used to determine the bias. 


\subsection{Data Processing}

Initial image processing of the \magixs\ data includes bias subtraction, dark current subtraction, gain adjustment, bad pixel removal, and despiking. \\

\noindent \textbf{Bias:}  The non-active regions of the images are used to determine the bias pedestal, which varies slightly between quadrants and as a function of time.  The bias is calculated and removed per frame and per quadrant. \\

\noindent \textbf{Dark Current:}  During ascent and descent, dark frames are obtained matching the exposure time of the science images (i.e. the images exposed to sunlight).  Due to a continual temperature increase in the analog chain during data collection that causes an increase in dark current, a pre-master dark is created from dark frames on the ascent and a post-master dark is created from dark frames on the descent.  The same number of dark frames are used to create the master darks.  The averaged analog chain temperature is stored with the associated master dark.  To deal with radiation hits, any pixel greater than 3 times the standard deviation of the pixel over the darks is ignored before creating the master dark.  Dark current is removed from a science image by creating an interpolation between the pre- and post-master dark using the science image's analog chain temperature. \\

\noindent \textbf{Gain:} The gain was measured with a Fe-55 source during pre-flight testing of the \magixs\ camera to be 2.6 electrons per Data Number (DN).   \\

\noindent \textbf{Bad Pixels:} Bad pixel maps are created using dark images.  Bad pixels are evaluated for each pixel location over the set of dark images.  If the values are not within 3 times the standard deviation of the mean for at least 80 percent of the time, the pixel location is marked as bad. Bad pixels are replaced by taking the median of the surrounding pixels.  The replaced locations and values are stored in a table with the image. \\

\noindent \textbf{Despiking:} Despiking uses a recursive technique to replace pixels of suspected radiation hits.  A list is created of pixels over a specified threshold.  Pixels with values from lowest to the highest are evaluated to allow for subtraction of radiation hits.  Images are divided into areas where individual thresholds can be applied.  If a pixel is determined to be replaced, the median of the surrounding pixels is used.  The replaced locations and values are stored in a table with the image. \\

Data sets have been generated for varying levels of processing.  Level~0.1 is the bad pixel mask.  Level~0.2 contains the pre- and post-master darks.  Level~0.5 is the Level~0 image set with timestamp adjustment applied, image acquisition state defined, and the CCD holder and cold block temperatures included.  The image acquisition states are dark, pointing, light, shutter door opening, and shutter door closing.  Level~1.0 is processed from the Level~0.5 sun-exposed (``light") images with the the bias and dark subtracted, the gain adjusted, and the bad pixels corrected.  Level~1.5 is the despiked Level~1.0 image set.


\subsection{Flight Calibration}

\begin{figure}
    \centering
    \includegraphics[width=1\textwidth]{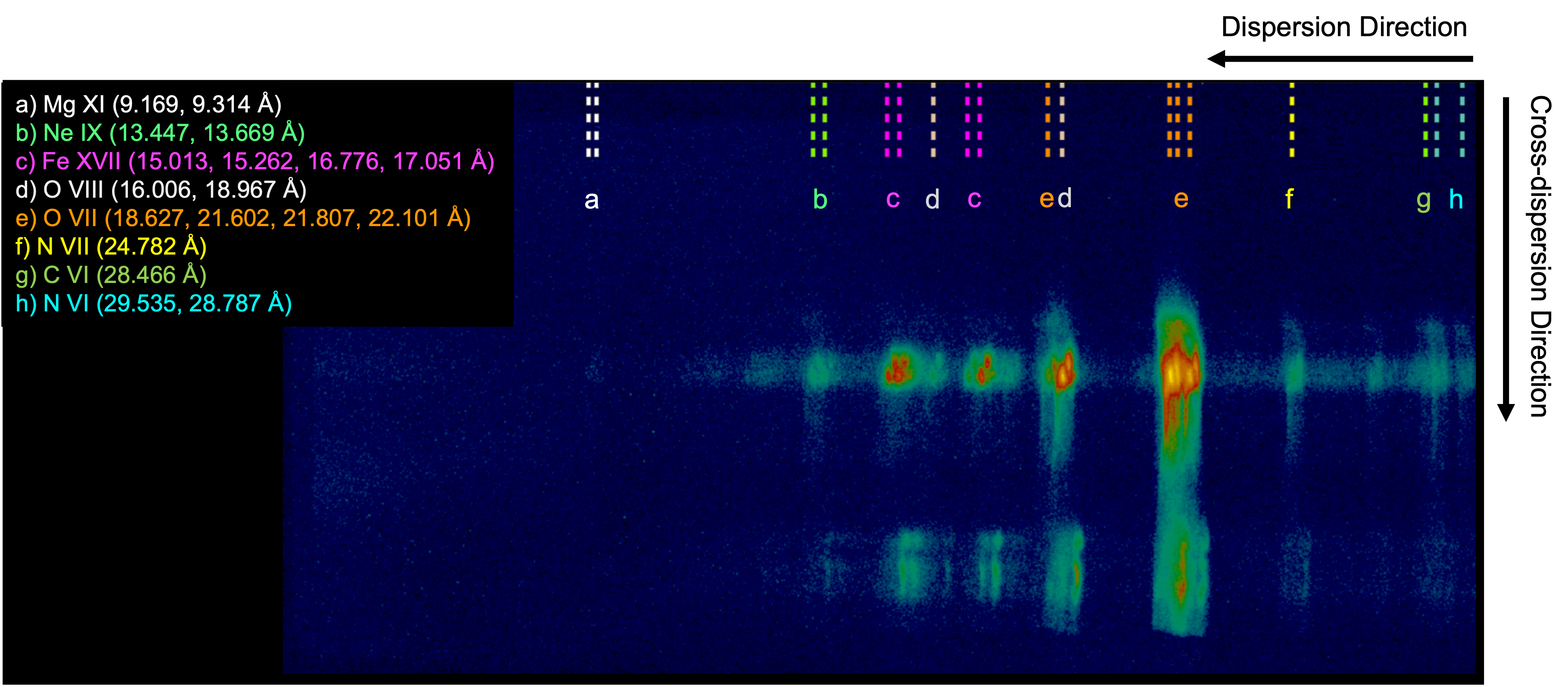}
    \caption{Level~1.5 summed \magixs\ data over the entire flight with key spectral line positions indicated along the dispersion direction.}
    \label{fig:flightdata}
\end{figure}

For flight calibration, we considered Level~1.5 processed images from stable pointing, which was established 185 seconds after launch (see Table~\ref{tab:pointing}). Figure~\ref{fig:flightdata} shows \magixs\ Level~1.5 data summed over 148 frames, which is used for flight calibration. 

\subsubsection{Spatial Plate Scale}

\begin{figure}
    \centering
    \includegraphics[width=0.9\textwidth]{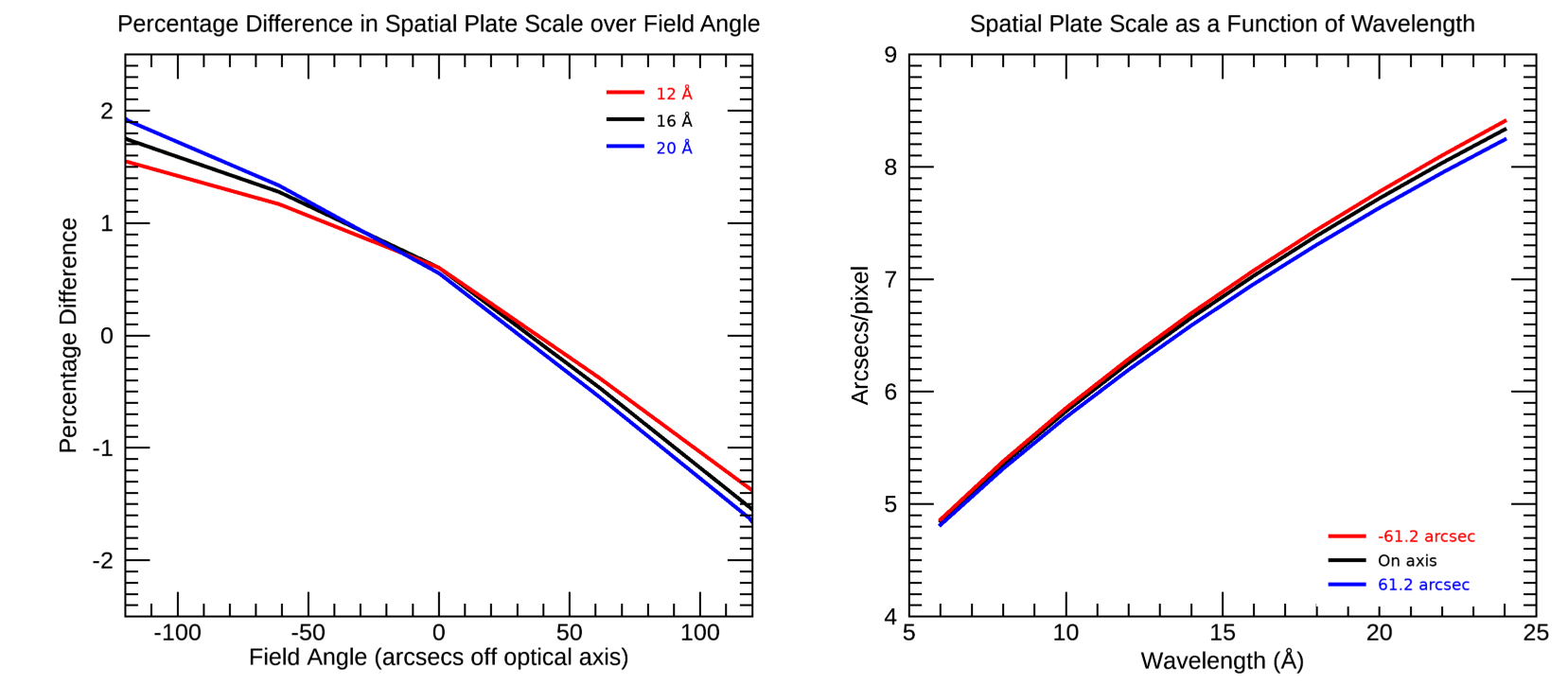}
    \caption{(Left) The spatial plate scale at 16\,\AA\ as a function of the field angle from the optical axis.  (Right) The plate scale averaged over field angle as a function of wavelength.}
    \label{fig:spatial_scale}
\end{figure}

In the cross dispersion direction, the spatial plate scale is 2.8{\arcsec}/pixel.  In the dispersion direction, the spatial plate scale varies with both field angle and wavelength, which is a property of the variable spacing reflective grating used in \magixs.  These relationships are demonstrated in Figure~\ref{fig:spatial_scale}, derived using an optical model of the grating in Zemax.  A table of the average spatial plate scale for key spectral lines is given in Table~\ref{tab:key_spec_lines}. 

\begin{deluxetable}{ccc}
\tabletypesize{\small}
\tablewidth{6in}
\tablecolumns{3}
\tablecaption{Key spectral lines and corresponding spectral plate scale values. \label{tab:key_spec_lines}}
\tablehead{\colhead{Ion}& \colhead{Wavelength} & \colhead{Average Plate Scale}\\
\colhead{}& \colhead{[\AA]} & \colhead{[\arcsec/pixel]}}
\startdata
N VI & 29.535   & 9.21\\
N VI & 28.787   & 9.09\\
C VI & 28.466   & 9.04\\
N VII & 24.782 & 8.45\\
O VII & 22.101 & 8.01\\
O VII & 21.602 & 7.93\\
O VIII & 18.967 & 7.51\\
O VII & 18.627 & 7.45\\
Fe XVII & 17.051 & 7.20\\ 
Fe XVII & 16.776 & 7.15\\
O VIII & 16.006 & 7.00\\
Fe XVII & 15.262 & 6.86\\
Fe XVII & 15.211 & 6.85\\
Fe XVII & 15.013 & 6.81\\
Ne IX & 13.699 & 6.56\\
Ne IX & 13.447 & 6.51\\
Fe XVII & 12.124 & 6.25\\
Mg XI & 9.314 & 5.60\\
Mg XI & 9.169 & 5.57\\
\enddata
\end{deluxetable}

\subsubsection{Wavelength Calibration}

One critical aspect of analyzing \magixs\ data is to  determine the wavelength calibration, meaning the wavelength as a function of pixel value. From ground calibration (see \citealt{Athiray2021}), we know the relationship is non-linear and varies as a function of field angle, meaning photons from different locations on the Sun experience different dispersion. Despite completing significant pre-flight calibration at the X-Ray \& Cryogenic Facility (XRCF) at MSFC \citep{Athiray2021}, only the wavelength calibration for a single field angle (``on-axis") was determined.  Additionally, the wavelength calibration shifted between the measurements taken at the XRCF and during flight, either due to pre-flight vibration or 1-g offloading.  Hence, the wavelength calibration as a function of field angle must be determined from flight data. 

\begin{figure}
    \includegraphics[width=0.60\textwidth]{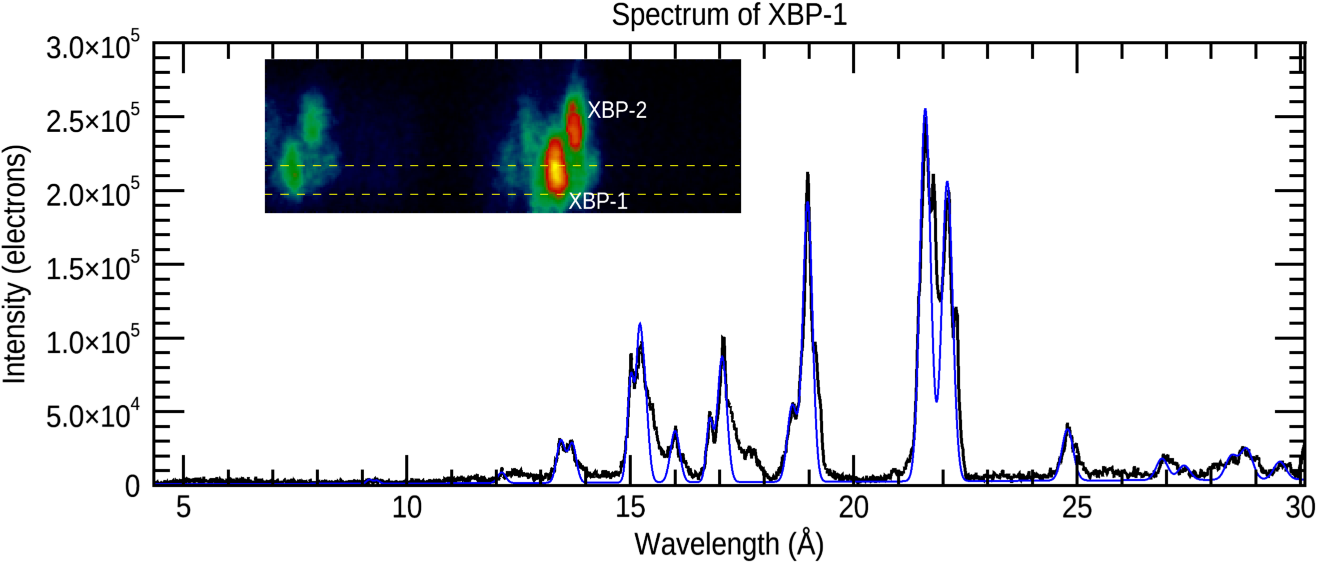}
    \includegraphics[width=0.4\textwidth]{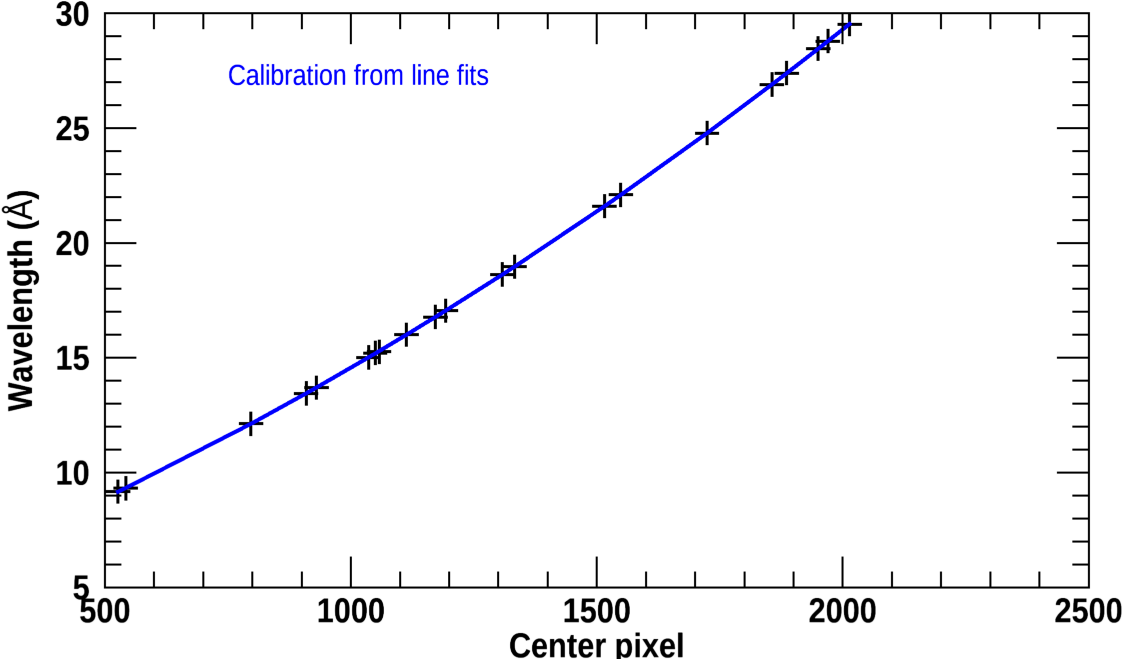}
    \caption{(Left) Spectral profile of XBP-1 derived from slices along the dispersion direction, summed over several cross-dispersion rows. The black curve represents the flight data, while the blue curve is the fit to the bright lines used to determine the centroids. (Right) Wavelength calibration function derived by mapping the centroid positions of the bright XBP-1 lines with position on the detector.}
    \label{fig:wavelength calibration}
\end{figure}

We considered a slice of the \magixs\ spectrum, summed over several rows along the cross-dispersion direction sampling the X-ray bright point-1 (XBP-1) in the northern targeted active region (refer to Figure~\ref{fig:target}). The resulting spectrum is shown in Figure~\ref{fig:wavelength calibration} (left). We identified prominent emission lines in the spectrum, modeled with a Gaussian function to derive respective centroid pixel locations.  We then performed wavelength calibration. Figure~\ref{fig:wavelength calibration} (right) shows the wavelength calibration for XBP-1, which is best modeled using a quadratic fit.  We define XBP-1 as our reference point and designate it with 0\textdegree\ field angle. This assumption implies the derived wavelength calibration is applicable for the ``effective on-axis'' (i.e., 0\textdegree) field angle. Using these coefficients, we then create a map of wavelength arrays for different field angles using the squashing and spectral plate scale derived from Zemax optical model. 

To verify and validate wavelength calibration and the map of wavelength array per field angle, we compared the spatial distance between XBP-1 and XBP-2 from \aia\ and \magixs\ images. The spatial separation in the x-direction between XBP-1 and XBP-2 in the rolled solar coordinates determined from the AIA\,335\,\AA\ image is $\approx$\,78\arcsec. This value closely matches and is consistent with the distance measured using the flight data ($\sim$\,70\,--\,78\arcsec), which confirms that the assumed squashing factor derived from Zemax optical model agrees with the derived wavelength calibration. 


\subsubsection{\label{subsec:rollpointing}Roll, Pointing, and Vignetting Function}

\begin{figure}
    \centering
    \includegraphics[width=0.4\textwidth]{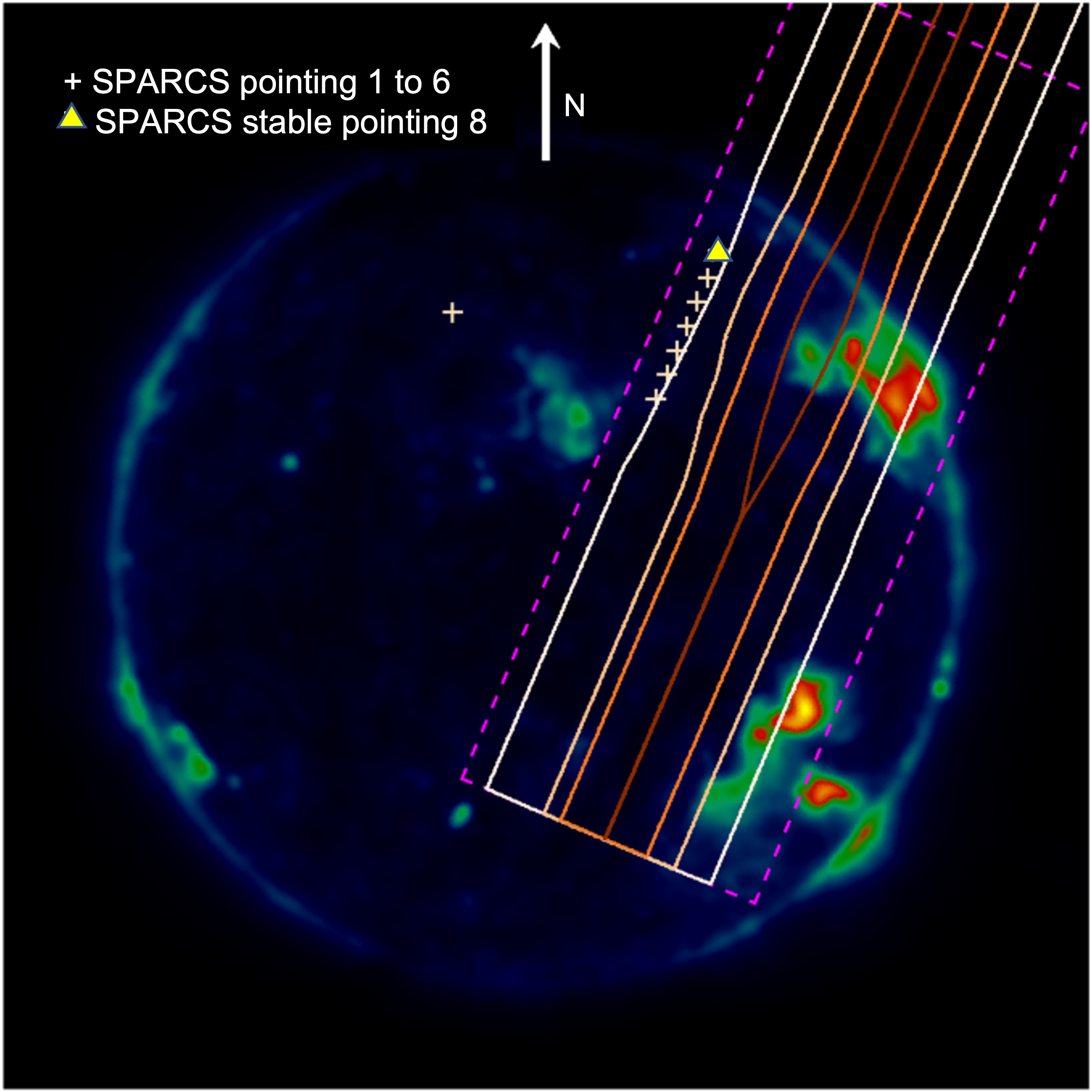}
     \includegraphics[width=0.47\textwidth]{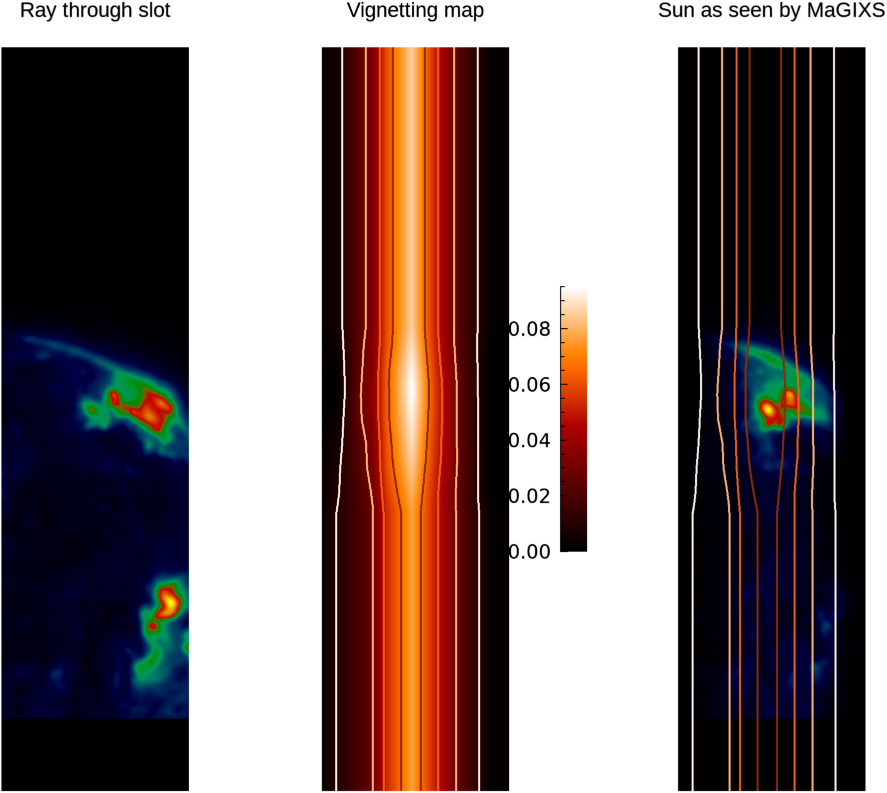}
    \caption{(Left) Vignetting contours (solid) overlaid onto an \aia\,335\AA\ image.  The dashed box indicates the extent of the physical slot.  The crosses indicate the the SPARC pointings. (Middle) Derived contoured vignetting map. (Right) Effective slot seen by \magixs\ due to the slot misalignment and optical vignetting.}
    \label{fig:rollpointing}
\end{figure}

The roll angle was determined 
from pointing adjustments made during the first 80 seconds of flight observation, where the SPARCS was commanded to point to several targets on the Sun such that the instrument was moved along the cross-dispersion direction for 54~seconds. Table~\ref{tab:pointing} lists the SPARCS pointing coordinates from Sun center, which are marked on the full disk AIA\,335\,\AA\ image shown in Figure~\ref{fig:rollpointing} (left). Using the relative offset between these SPARCS pointings, we determined the roll angle to be 23$^{\circ}$ clockwise about solar North. 
 
During the ground tests, the Lockheed Intermediate Sun Sensor (LISS), an element in SPARCS used for fine pointing, and \magixs\ Wolter-I telescope were co-aligned to within 1\arcmin\ accuracy using a theodoloite and an auto-collimator. Despite significant efforts to co-align the \magixs\ optical surfaces, it was discovered during the pre-launch heliostat test at White Sands, just two weeks before launch, that the Wolter-I and the center of the slot were misaligned from the LISS. The measured offset of slot center from LISS, determined from the heliostat test was $\sim$6\arcmin.5\,$\pm$\,1\arcmin, which was factored into the pre-flight pointing calculations. 

The offset between the LISS and the alignment of the TMA and SOA could not be accurately determined from the heliostat tests. However, flight data indicates that an offset between the LISS pointing, slot center, and the optical surfaces introduced additional vignetting, which is non-trivial to model. 
Therefore, we use the flight data combined with optical models with variable offset scenarios to determine appropriate vignetting maps that correspond to the flight observations. 

To determine the absolute pointing post-flight, we forward modeled \magixs\ observations using a differential emission measure (DEM) map derived from time-averaged coordinated {\aia} observations (channels used: 94, 131, 171, 193, 211 and 335\,\AA), the \magixs\ instrument response, and the vignetting maps for different offset scenarios. We selected pixels around the brightest O~VIII emission line at 18.967\,\AA, used as our reference to compare flight and forward models. The selected \magixs\ image was co-aligned with a forward model through cross-correlation techniques, thus determining the optimal offset between the LISS and slot center and defining a plausible vignetting function. Figure~\ref{fig:rollpointing} shows the most likely absolute solar pointing overlaid with the vignetting map (marked as contours), which forms the `effective slot' (9.2{\arcmin} x 25{\arcmin}). 
Figure \ref{fig:rollpointing} (right) shows the portion of the Sun that reached the \magixs\ grating and science camera.  

Consequently, this first \magixs\ flight missed making observations of the brightest portions of the target active regions due to the significant impact of the slot offset and internal vignetting.  Yet, the measurements taken from the captured portions of the two active regions successfully yielded X-ray spectra, a monumental feat in itself.  These \magixs\ data contain a wealth of information, prove the performance of the design, and provide the opportunity to optimize the inversion techniques for future observations.
 

\subsection{\label{inversion}Inversion of \magixs\ Data}

To unfold the \magixs\ data, we follow the general framework of spectral decomposition described in \citep{2015ApJ...807..143C,2019ApJ...882...13C,2019ApJ...882...12W}, and described briefly here.  We first cast the problem as a set of linear equations, namely
\begin{equation}
    y = Mx
    \label{eqn:lin}
\end{equation}
where $y$ is an array that contains a single row of the \magixs\ flight data, $M$ is a matrix that contains how emission at each solar location and temperature map into the detector, and $x$ is an array of emission measures at different solar locations and temperatures.  The \magixs\ response matrix, $M$, is generated  using the wavelength calibration as a function of field angle determined from flight data and effective area measured pre-flight \citep{Athiray2021}.  Using the CHIANTI atomic database v10.0.2 \citep{chianti_v10}, we construct several variants of isothermal, unit EM instrument response functions with different abundances (i.e., coronal/photospheric), different electron distributions (i.e., Maxwellian/kappa), and an assumption of ionization equilibrium.  Equation~\ref{eqn:lin} is then solved for $x$ using the ElasticNet routine in SciPy, a Python library. ElasticNet allows for varying the extent of smoothness and sparseness while finding convergence to the best solution.  (Using ElasticNet is a difference from the previous published papers, which used the LassoLars routine.)  

The ElasticNet routine is solving Equation~\ref{eqn:lin} by finding:
\begin{equation}
    x^{\#} = argmin\left[ ||y-Mx||^2_{2} + \alpha\rho||x||_{1} + 0.5\alpha(1-\rho)||x||^2_2\right]
    \label{eq:soln}
\end{equation}
where $\alpha$ is the penalty term and $\rho$ is varied from 0 to 1.  The first term in Equation \ref{eq:soln} is the standard least squares term that minimizes the difference between the observations and the forward calculated observations.  The second term is the $L_1$ norm of $x$, minimizing this term favors a sparse solution.  The third term is the $L_2$ norm of $x$, minimizing this term favors a smooth solution.  Increasing $\alpha$ increases the weight of the penalty.  For $\rho = 1$, the solution will be sparse, while for $\rho = 0$, the solution will be smooth.  We inverted the data with a variety of $\alpha$ and $\rho$ solutions, comparing how well the inverted data, $Mx$, matched the observations, $y$.  We also considered different inversion routines that were purely sparse (such as LassoLars) or purely smooth.  We will present the parameter space search and provide a quantitative comparison of different algorithms in a future paper.  For the Level 2 data included in the initial analysis below, we use $\alpha = 1\times 10^{-5}$ and $\rho = 0.1$.

\begin{figure}[!h]
    \centering
    \includegraphics[width=0.8\textwidth]{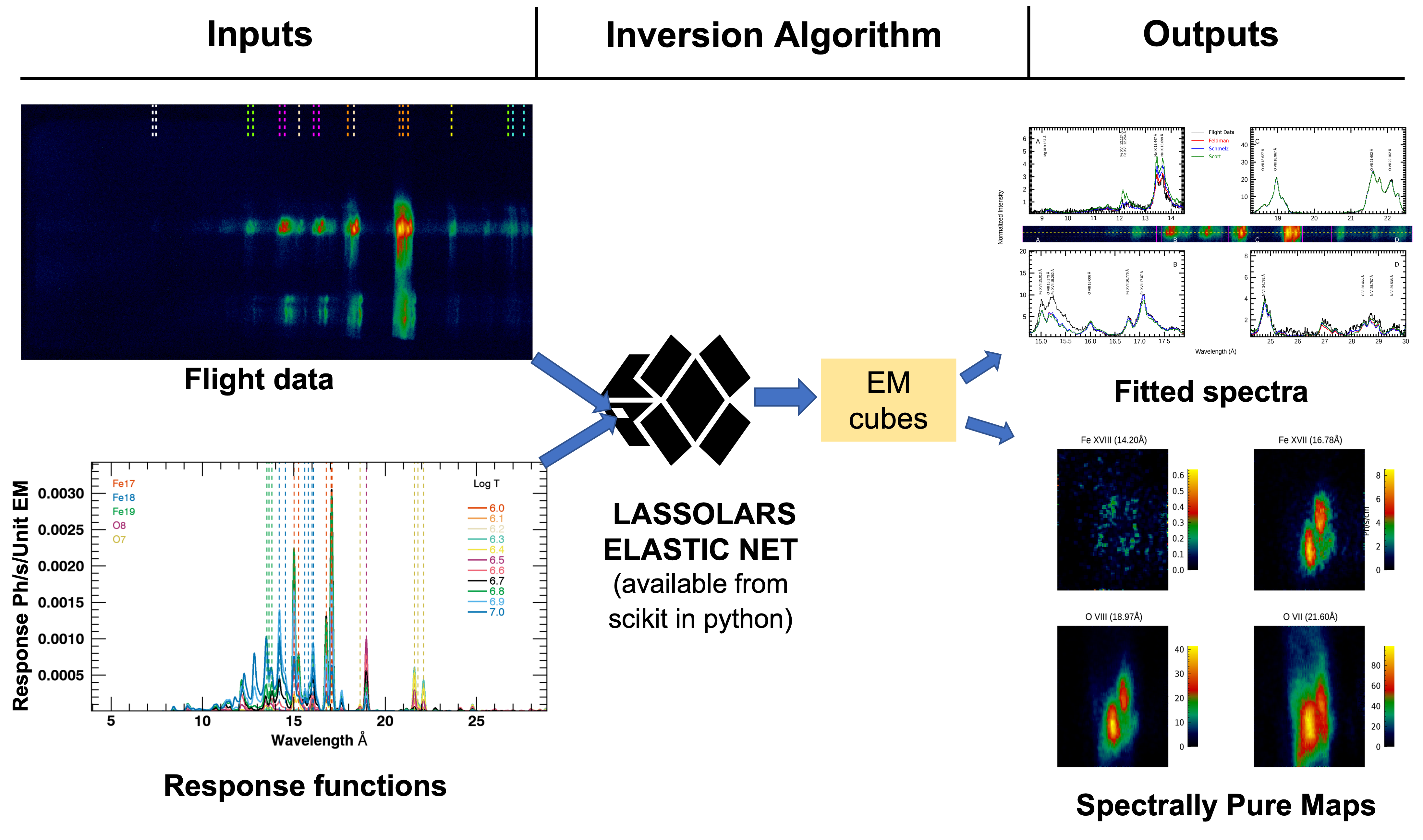}
    \caption{Schematic description of the inversion process of converting a spectroheliogram to pure spectral maps.}
    \label{fig:inversionprocess}
\end{figure}

\begin{figure}
    \includegraphics[width=1\textwidth]{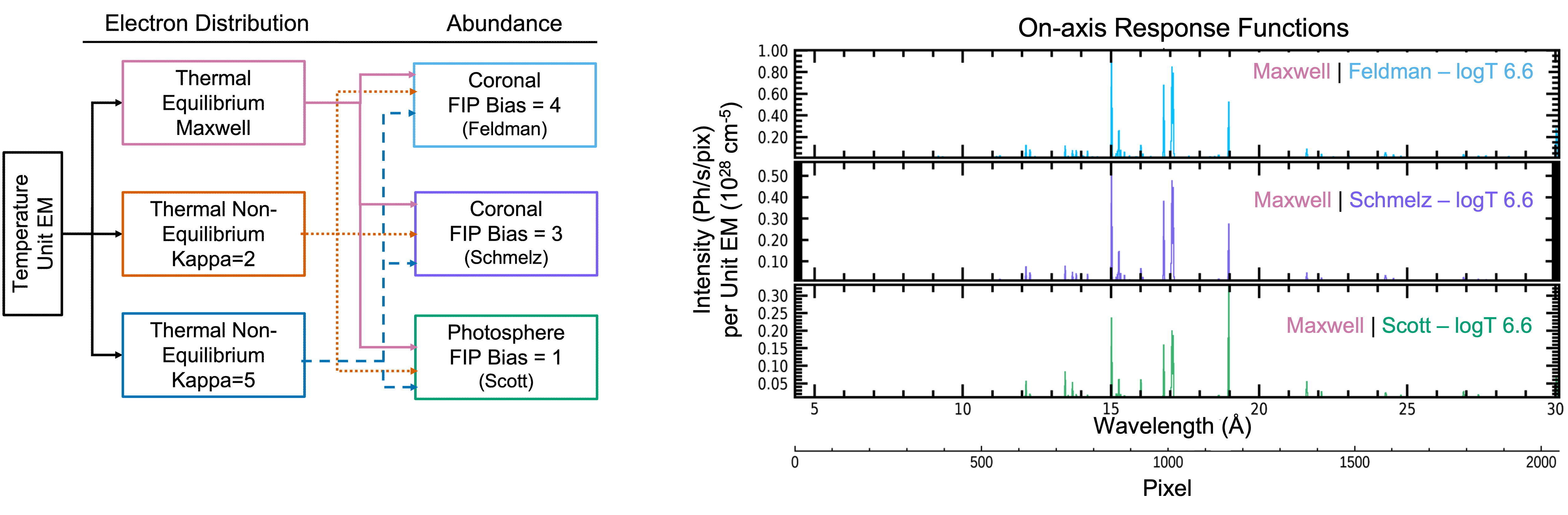}
    \caption{(Left) Depiction of atomic parameter space applied to the generation of the response functions. (Right) Sample response functions for an on-axis source at plasma temperature logT=6.6. The three panels correspond to response functions for different abundances under Maxwell electron distribution. All of the response functions are generated under the assumption of ionization equilibrium. }
    \label{fig:resp_combination}
\end{figure}

Figure \ref{fig:inversionprocess} provides a high-level schematic of the inversion process, wherein the instrument response function and spectroheliogram data are inputs and emission measure cubes, fit spectra, and spectrally pure maps of different ion species are the outputs of the inversion. We perform several inversion runs with different response functions as a varying input. In addition to the field angle-wavelength map and effective area, the response function invokes a combination of atomic parameters involving temperature, electron distribution, and abundances as listed in Figure \ref{fig:resp_combination}. All of the responses are constructed under the assumption of ionization equilibrium. Figure \ref{fig:resp_combination} (right) shows an example of response functions for an on-axis source for different abundances using a Maxwell electron distribution. We then determine multiple inverted solutions for various possible input combinations and compare the results with the flight fitted spectrum to select the best match solution. 

\subsubsection{Example Inversion}

\begin{figure}
    \centering
    \includegraphics[width=0.74\textwidth]{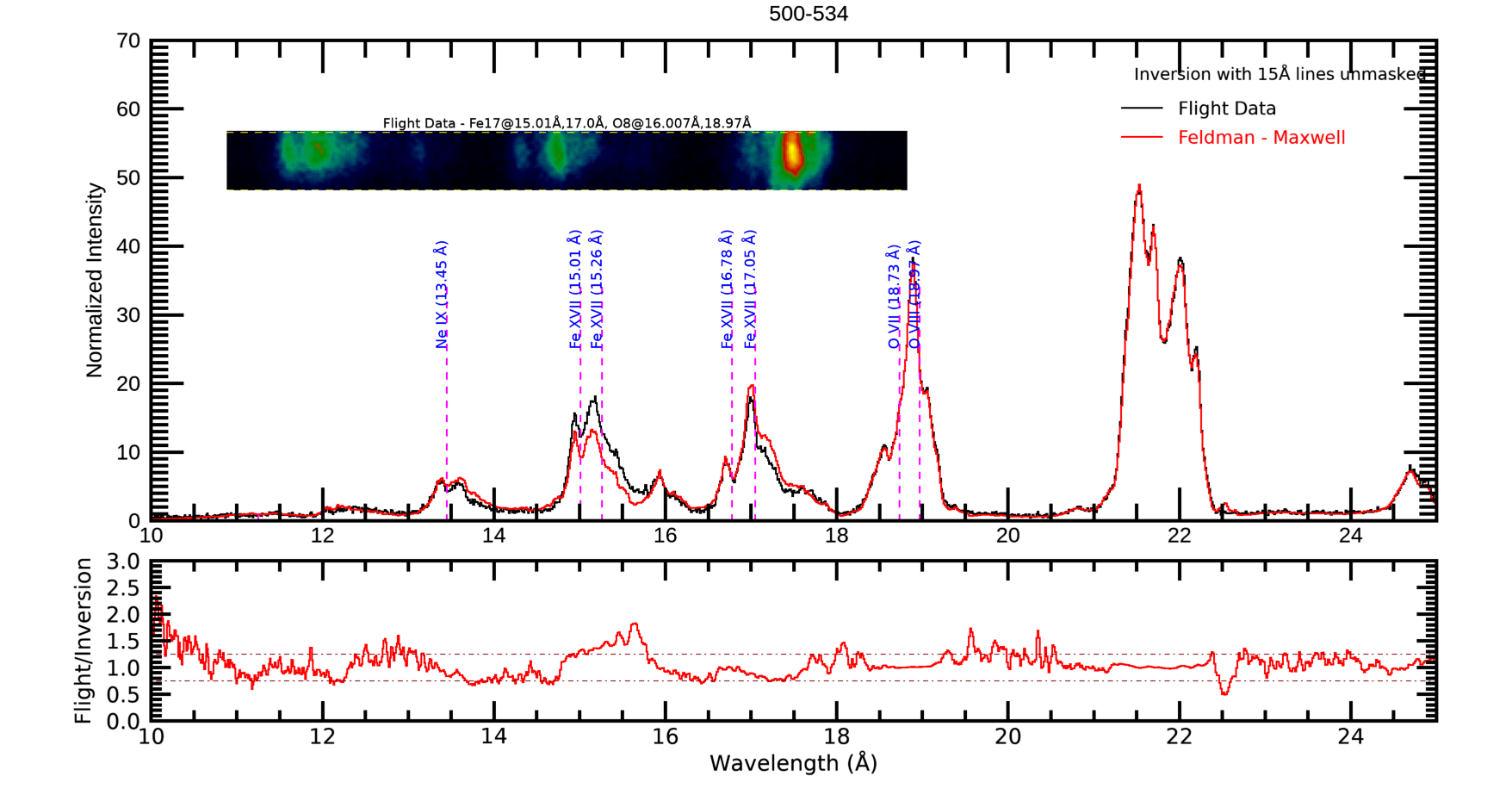}
    \caption{Comparison of flight and inverted spectra using coronal abundances \citep{Feldman92} for the XBP-1 region. The inset shows the spectroheliogram. Bottom panel shows the ratio of flight to inversion.}
    \label{fig:nomask_inversion}
\end{figure}

Figure~\ref{fig:nomask_inversion} provides an example of the inversion of the XBP-1 spectroheliogram.   The inset shows a cropped section of the \magixs\ data around the XBP-1 in the 15-20 \AA\ wavelegth range.  The spectrum from the selected region of XBP-1 is marked with dashed horizontal lines in the inset and plotted with a black line.  This sample inversion uses a Maxwell electron distribution, ionization equilibrium, and Feldman coronal abundances \citep{Feldman92}. The fit spectrum from the inversion results are shown as a red line.  The fit spectrum near Fe XVII 15\,\AA, including the collisional 3C line at 15.013\,\AA\ and the intercombination 3D line at 15.262\,\AA\ \textbf{do not agree with the observed flight data}. Specifically, the observed flight data shows an \textbf{excess emission} in the wavelength range of $\sim$\,15\,--\, 15.7\,\AA, which is {\bf consistently brighter than the results from the inversion, regardless of inversion parameters}. This discrepancy is consistent across all spatial structures (e.g., it is not simply limited to XBP-1). 


\subsubsection{Possible Explanations for Excess Emission}

We carried out a multi-pronged investigation to understand and explain this wavelength dependent discrepancy considering instrument artifacts and completeness of the atomic database in generating the instrument response.  The initial expectation was that the presence of absorption edges (near $\sim$14.9\,\AA) of the Ni and Ir used in X-ray mirror coatings could have impacted the shape and confidence in the derived effective area. A thorough study, however, indicated that to match the observed intensity around the $\sim$\,15\,--\,15.7\,\AA\ would require $\sim$60\% more effective area in this narrow wavelength band, while the current effective area matches well for all other emission lines, including the Fe~XVII line at 16.776\,\AA. We acknowledge lack of effective area measurements across the Ni edge, however we argue that a 60\% increase in the effective area in a narrow wavelength range is not physical considering that the remaining broad wavelength range agrees with the measured effective area curve (see \cite{Athiray2021}).

We performed a comparison between high-resolution solar spectra and CHIANTI data and found that there are several missing transitions in the database around 15\,\AA. From analyses of previous solar and laboratory spectra \citep{Beiersdorfer_2012, Beiersdorfer_2014, Lepson2017}, we know that satellite lines of Fe~XVII, Fe~XVI, and Fe~XV ions are present in this wavelength range. We  also know that some are missing in the database, hence they are not included in the response functions. 
It is well known that a historical discrepancy of the Fe~XVII 3C/3D line ratio arises due to the proximity of a satellite line of Fe~XVI (15.261\,\AA) to the Fe~XVII 3D line. Recent observations from both EBIT and astrophysical sources indicate that the Fe~XVI and Fe~XV ions emit a series of emission lines, including several satellite lines, which are blended or are very closely spaced with Fe~XVII lines, from 15.01\,\AA\ to 15.7\,\AA\ \citep{Graf_2009, Brown2008}. 

Generally, in active regions and flares, the satellite lines are much weaker than the Fe~XVII lines. The atomic data for the strong Fe~XVII lines indicate agreement within 10\% with solar high-resolution spectra \citep{delzanna:2011_fe_17},
hence the problem cannot be with the atomic data for this ion.

\begin{figure}
\centering
    \includegraphics[width=0.4\textwidth]{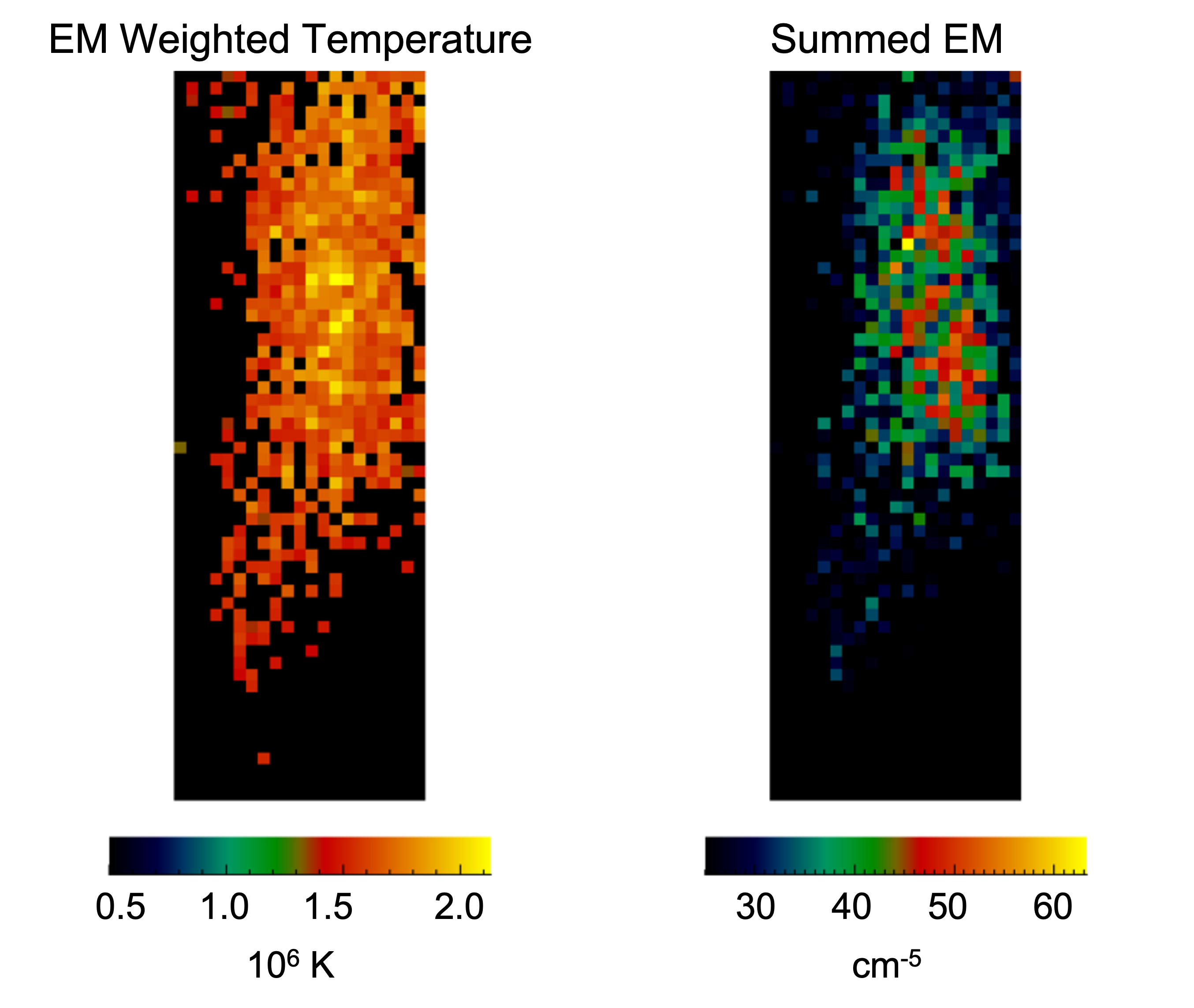}
    \caption{(Left) Emission measure weighted temperature map for XBP-1. (Right) Corresponding summed EM cube.}
    \label{fig:EM_weighted_temp}
\end{figure}

Fe~XVI is the simplest ion with only one valence electron in the M shell; Fe~XV is the next simplest ion with two valence electrons in the M shell. These ions exhibit peak emissivity near 2\,MK. The EM-weighted temperature for the XBP-1 region calculated from an inverted EM cube (see Figure \ref{fig:EM_weighted_temp}) indicates that 2\,MK, `relatively cool' plasma emission is dominant.  Therefore, the region is expected to have brighter Fe~XVI satellite lines than Fe~XVII, thus dominating/contaminating the spectral band around 15\,\AA. 

The missing flux due to satellite lines would therefore have a significant effect and could explain the discrepancy. Work is in progress to calculate the atomic data and update the CHIANTI database. In the meantime, given the incompleteness of the atomic data around 15~\AA\ for such low temperatures, we have chosen to remove the 14.5\,--\,15.5\,\AA\ region from the analysis.

\subsubsection{De-convolved Spectral Maps}

Plasma diagnostics such as temperature, density, abundances, electron distribution can be derived from spectroscopic observations by measuring the line intensities and ratios of different spectral lines from different ion species at different ionization states. However, absolute line intensities cannot be directly deduced from spectroheliogram data of an extended source, such as an active regions or X-ray bright point, due to overlapping spatial-spectral information and therefore need to be inverted to yield spectrally pure maps (as described above in Section~\ref{inversion}). 

\begin{figure}
    \includegraphics[width=0.5\textwidth]{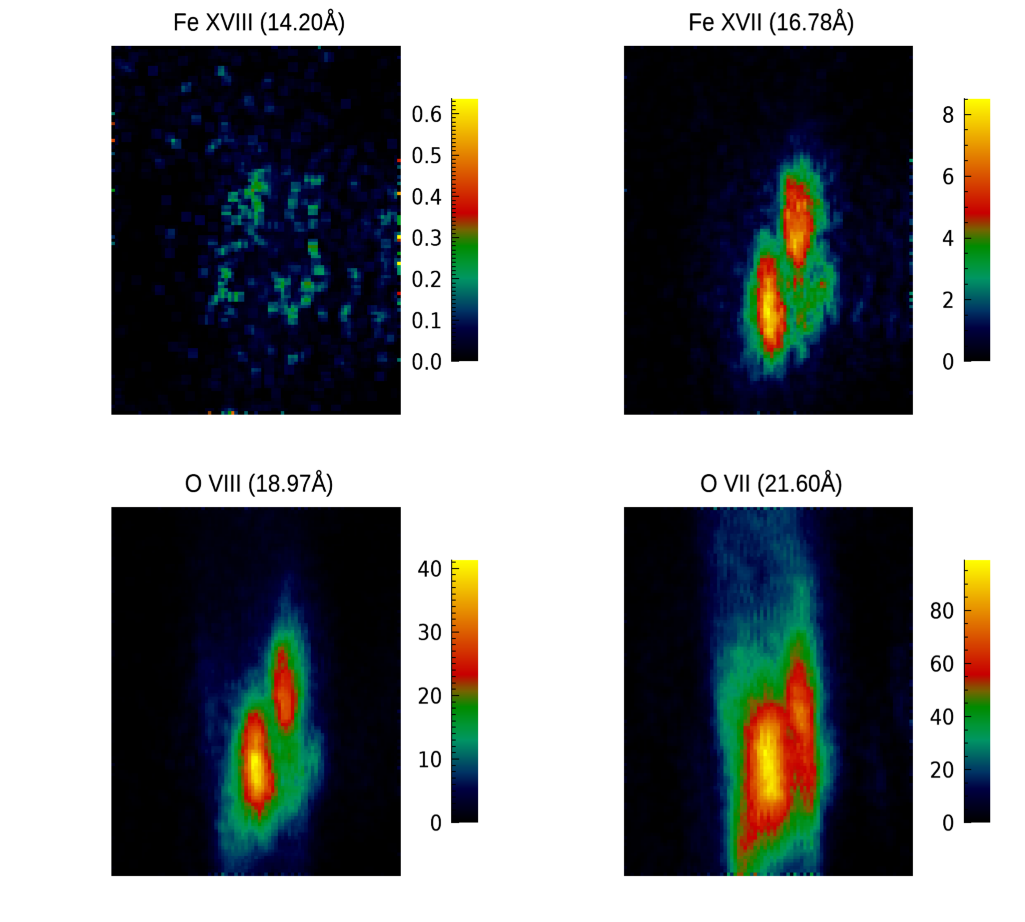}
    \includegraphics[width=0.5\textwidth]{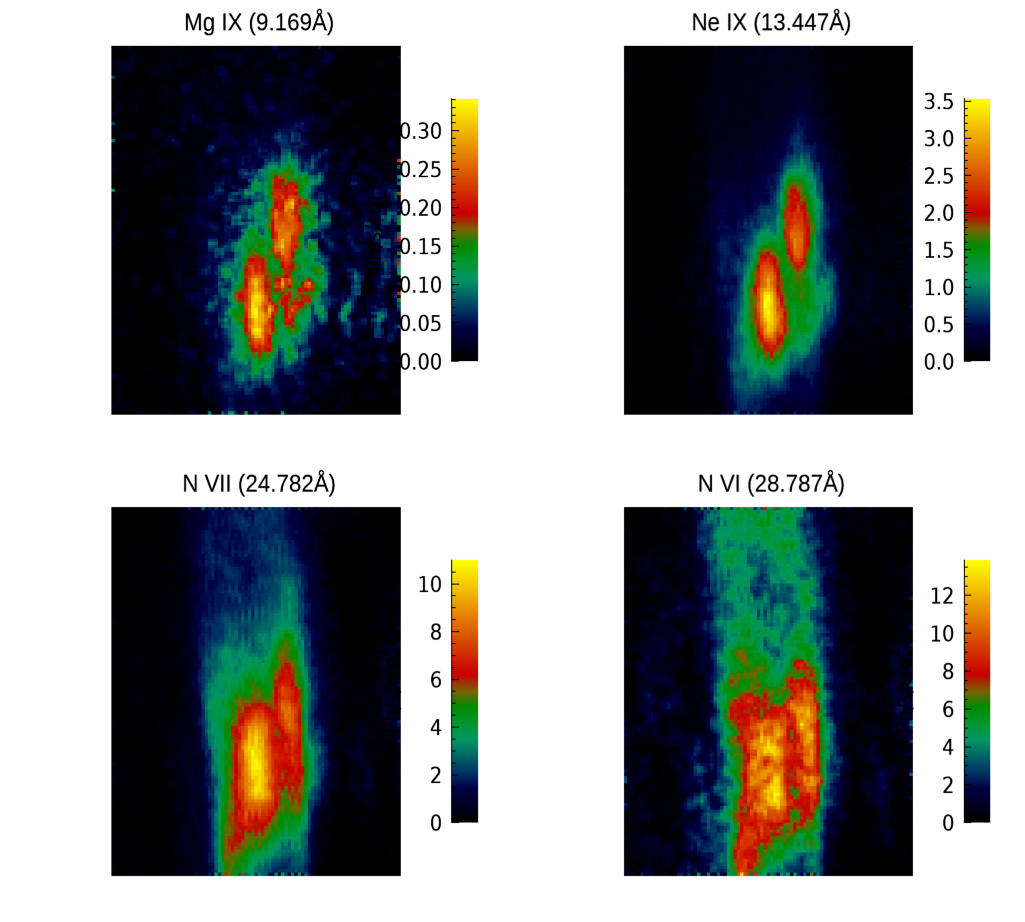}
    \caption{Spectrally pure maps of XBP-1 derived from inversion of the \magixs\ spectroheliogram in units of Ph/s/cm$^2$.}
    \label{fig:spec_pure_maps}
\end{figure}

\textbf{One of the primary \magixs\ data products derived from the inversion process is the generation of spectrally pure maps of different observed ion species for the full detector image.} These maps are obtained by folding the inverted emission measure cube through the emissivity functions of different ions created using CHIANTI database with the same atomic assumptions (abundances, ionization equilibrium, etc.). 
Figure~\ref{fig:spec_pure_maps} shows the spectrally pure maps of XBP-1 in Fe~XVII, Fe~XVIII, O~VIII, O~VII, N~VI, and N~VII.

\section{Coordinated Data}

To enhance the science return of the \magixs\ flight, coordinated data sets were specifically obtained from several external solar and astrophysical instruments.  These data sets are listed in Table~\ref{tab:coord-data}, along with continuously available \sdo\ data.  Note that while most of these data sets primarily targeted the brighter southern active region, spatially and temporally overlapping data corresponding to the \magixs\ field of view is available across the solar atmospheric layers.  Analyses targeting these coordinated sets will be the subject of forthcoming studies.

\begin{deluxetable}{ll}[h!]
\tabletypesize{\small}
\tablewidth{6in}
\tablecolumns{2}
\tablecaption{Available concurrent and complementary data sets observing at least one of the \magixs\ target regions. *IHOP 421: \href{ttps://www.isas.jaxa.jp/home/solar/hinode_op/hop.php?hop=0421} {https://www.isas.jaxa.jp/home/solar/hinode\_op/hop.php?hop=0421} 
\label{tab:coord-data}}
\tablehead{\colhead{Solar Target}&\colhead{Instrument} }
\startdata
\begin{tabular}[c]{@{}l@{}}Photosphere \\ \\ \end{tabular} &
  \begin{tabular}[c]{@{}l@{}}*\hinode\ Solar Optical Telescope (SOT)\\ Solar Dynamics Observatory (\sdo) Helioseismic and Magnetic Imager (HMI)\end{tabular} \\
   & \\
 \begin{tabular}[c]{@{}l@{}}Transition Region \\ \\ \end{tabular} &  
   \begin{tabular}[c]{@{}l@{}}\sdo\ Atmospheric Imaging Assembly (AIA) 1600 / 1700 \AA\\\ *Interface Region Imaging Spectrograph (\iris)\end{tabular} \\
   & \\
 \begin{tabular}[c]{@{}l@{}}Corona \\ \\ \\ \\ \end{tabular} &  
  \begin{tabular}[c]{@{}l@{}}\sdo\ AIA short-wavelength EUV Images\\ \hinode\ EUV Imaging Spectrometer (EIS)\\ \hinode\ X-Ray Telescope (XRT)\\ Nuclear Spectroscopic Telescope Array (\nustar)\end{tabular} \\* \midrule
\enddata
\end{deluxetable}


















\section{\label{discussion}Discussion}

Using the pure spectral maps provided in Figure~\ref{fig:spec_pure_maps}, we have performed targeted analysis of XBP-1. We consider four unique plasma diagnostics signatures, described in Section~\ref{introduction}, that contribute to the differentiation between the high- and low-frequency heating scenarios. \\



\begin{figure}[!h]
\center \includegraphics[width=0.5\textwidth]{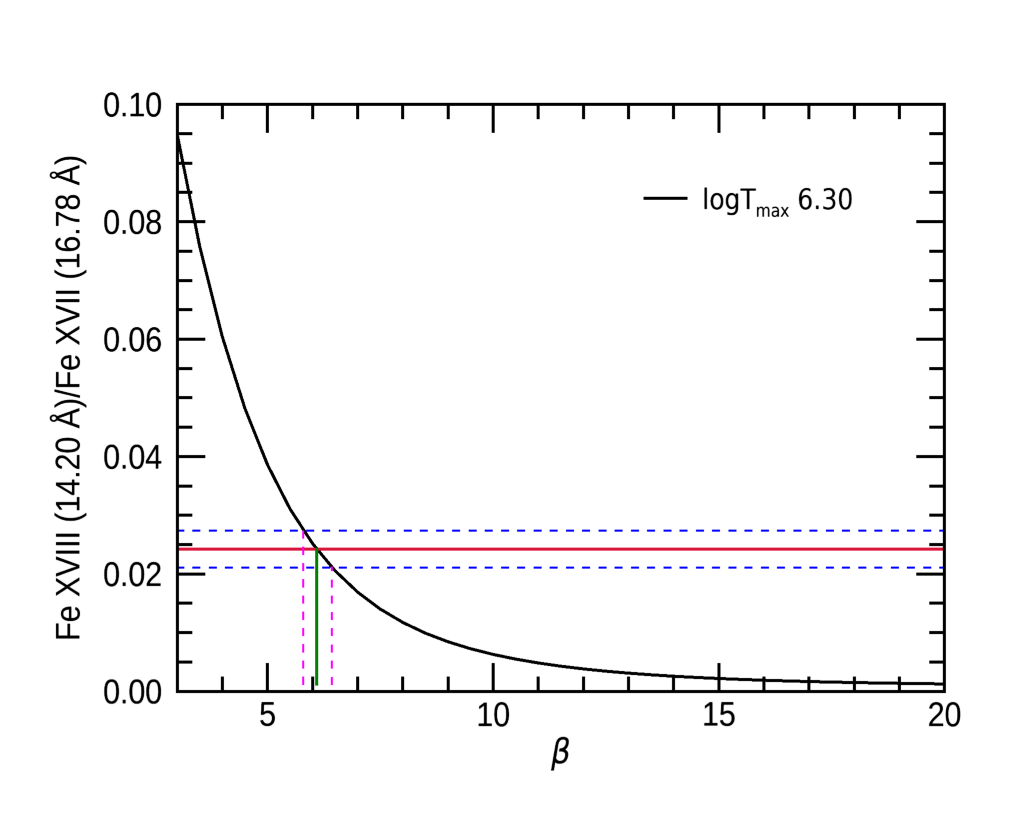}
    \caption{Intensity ratios of Fe~XVIII/Fe~XVII as a function of EM slope ($\beta$) with peak temperature at logT$_{max}$\,=\,6.30. With little/no dominant Fe~XVIII emission in \magixs, the observed ratio for the integrated XBP-1 region (0.024\,$\pm$0.003) serves as an \textbf{upper limit} for $\beta$ (denoted by the intersection of the solid red horizontal line with the solid green vertical line). The dashed vertical lines denote the range of allowed beta values corresponding to the uncertainty in the observed ratio.}
    \label{fig:beta_limit}
\end{figure}

\noindent \textbf{I. High-temperature, low-emission plasma:} Observing temperature sensitive diagnostic emission lines such as Fe~XVIII, Fe~XIX directly indicates the presence of hot plasma, and simple intensity ratios with Fe~XVII could help determine the heating frequency. \cite{athiray2019} showed that line ratios of Fe~XVIII and Fe~XVII can be directly related to the high temperature EM slopes ($\beta$). The XBP-1 region observed by \magixs\ distinctly emits Fe~XVII lines, while little/no significant Fe~XVIII emission is observed, as shown in the spectrally pure maps of XBP-1 in Fe XVII and Fe XVIII (Figure~\ref{fig:spec_pure_maps}). Therefore, we cannot strongly infer much about the high temperature fall off due to the lack of strong Fe~XVIII emission. However, we determine the ratio of Fe~XVIII to Fe~XVII using the total emission integrated over XBP-1, which sets an upper limit on the $\beta$ value for the entire XBP-1 region as shown in  Figure~\ref{fig:beta_limit}. The black solid line indicates the ratio of Fe~XVIII/Fe~XVII as a function of $\beta$ for a peak temperature at logT$_{max}$=6.30. The observed ratio with uncertainty (0.024\,$\pm$0.003) is denoted by horizontal lines (solid red, dashed blue). The intersection of the horizontal (red) and vertical lines (green) on the plot denote the upper limit for $\beta$ to the ratio derived from \magixs. The range of allowed $\beta$ values corresponding to the uncertainty in the observed ratio is represented by vertical dashed lines. The low ratio (0.024) suggests high $\beta$ (6.09), \textbf{indicating a high-frequency heating scenario for this bright point}.\\



\begin{figure}
    \centering
    \includegraphics[width=0.9\textwidth]{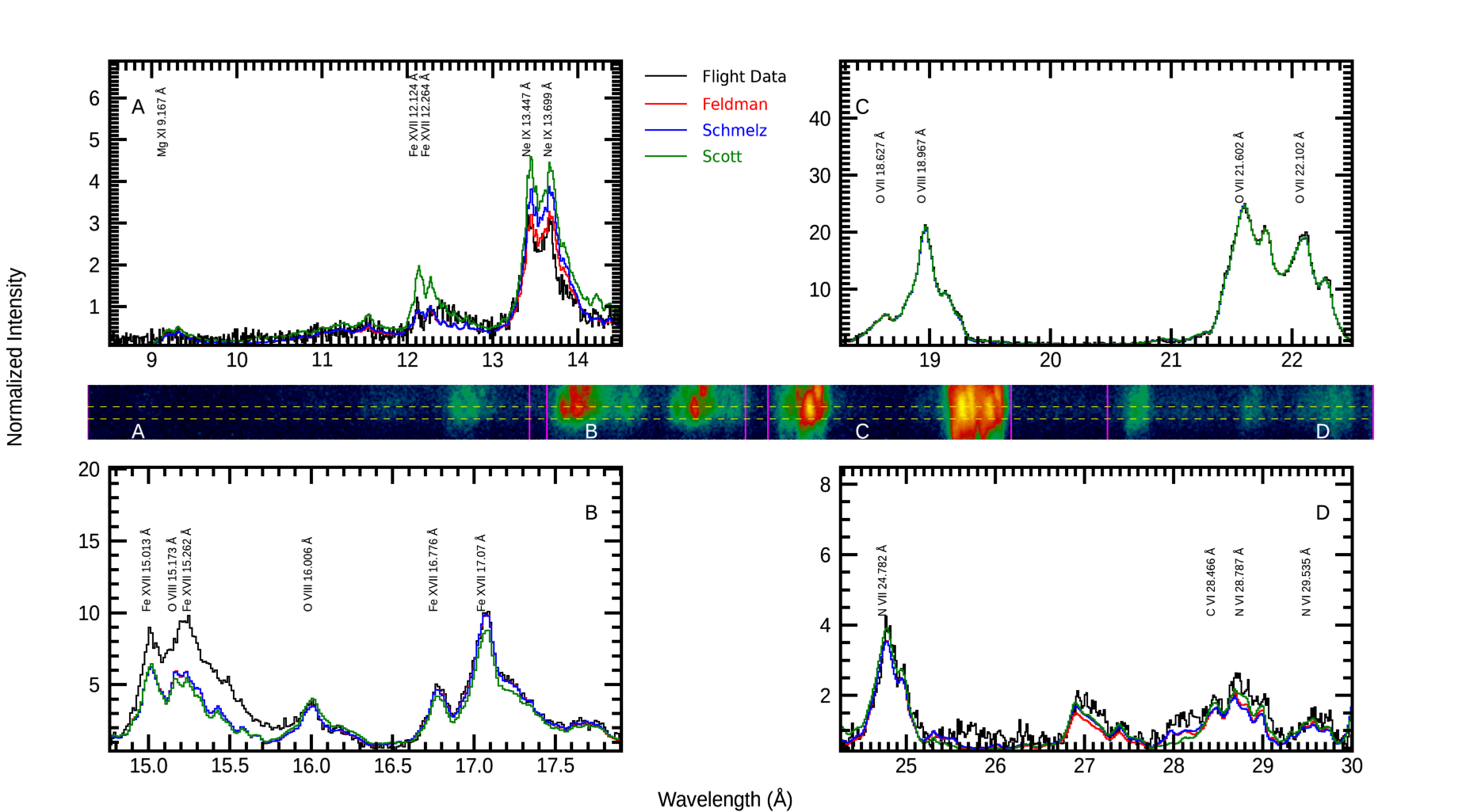}\caption{Comparison of flight and inverted spectra using different solar abundances (Coronal - \citet{Feldman92,schmelz2012}; Photospheric - \citet{scott2015}) for the XBP-1 region. Top and bottom panels (A, B, C, and D) compare spectra from different wavelength ranges with the prominent emission lines labeled. The cropped flight spectroheliogram is shown in the middle with the corresponding wavelength ranges marked with vertical lines. Horizontal dashed lines indicate the rows that are summed for the XBP-1 spectra.  Note all inversions underestimate the observed emission in the 15-15.7\AA\ wavelength range, as discussed in Section 4.4.2.}
    \label{fig:compare_abundances}
\end{figure}

\noindent \textbf{II. Element abundances of high temperature plasma:} Determining the abundances of elements in the solar corona is a primary objectives of the \magixs\ instrument. We expected from previous analyses for quiescent active region cores to have a FIP bias of 3\,--\,4 at the temperatures of the \magixs\ lines [\citep[e.g.,][]{2014A&A...565A..14D}, also refer back to Section~\ref{introduction} (II)].  However, recent line-to-continuum measurements reported by \cite{vadawale_etal:2021} of the full-Sun in X-rays have indicated that XBPs have a lower FIP bias, although those measurements were performed with a disk-integrated instrument.  To determine which abundance matches the \magixs\ data, we generated several \magixs\ response functions using coronal \citep{Feldman92, schmelz2012} and photospheric abundances \citep{scott2015}, as described in Section~\ref{inversion}. Inversions are performed using different response functions, and the results are compared in Figure~\ref{fig:compare_abundances}. The clearest abundance diagnostics in the relatively cool XBP-1 are from the Ne~IX lines between 13\,--\,14\,\AA, shown in panel A of Figure~\ref{fig:compare_abundances}.  Specifically, \textbf{coronal FIP bias ($\sim$4)} from \citet{Feldman92} \textbf{agrees most closely with the flight data}.  Note that the observed Ne lines are reproduced consistently by the Feldman model, whereas the Schmelz model overpredicts the measurements.   


\vspace{0.3in}

\begin{figure}
\center \includegraphics[width=0.5\textwidth]{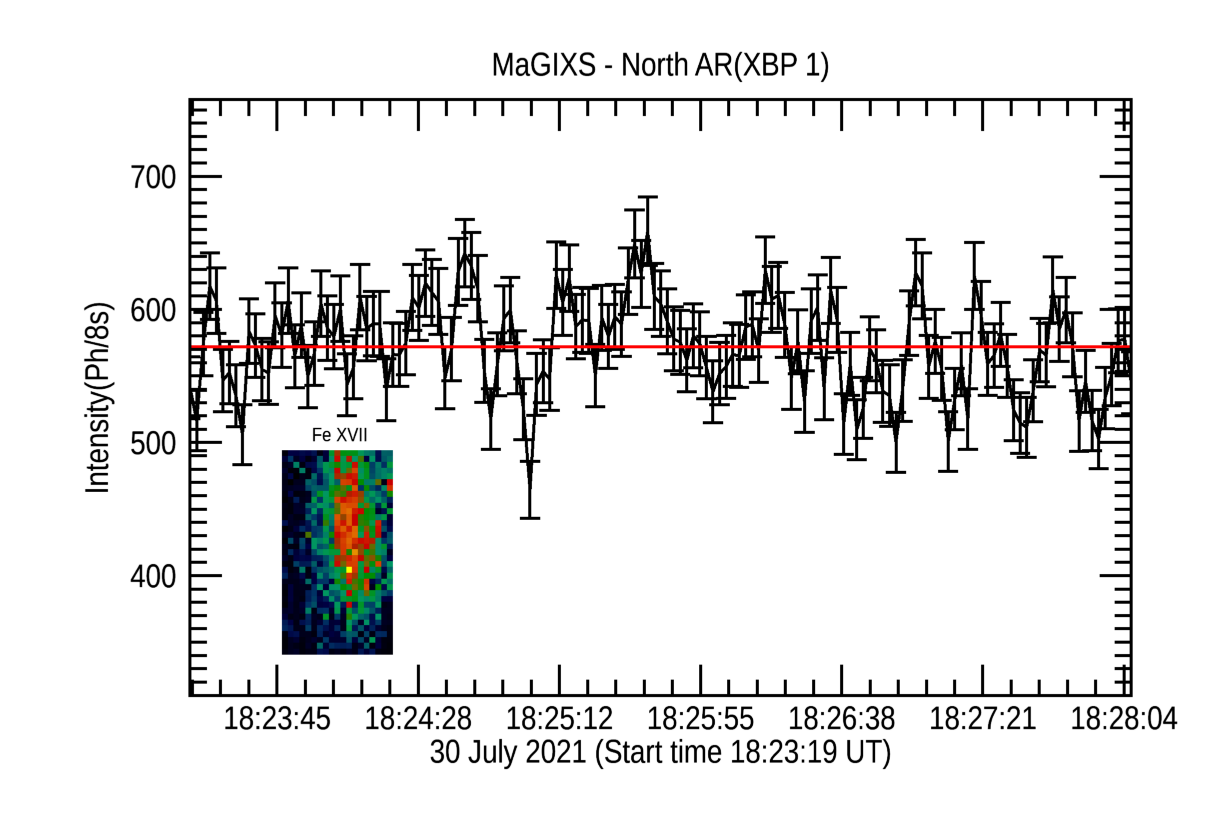}
    \caption{Light curve of Fe~XVII emission from XBP-1 constructed by inverting the running averaged data with four frames added at each time bin.  Error bars are from photon noise only and are 1$\sigma$.  Red line is the average intensity. }
    \label{fig:time_evolution}
\end{figure}

\noindent \textbf{III. Temporal variability of high temperature plasma:} The fluctuations of Fe~XVII line intensities with time probes the impulsiveness of heating events. The emissivity of Fe~XVII peaks $\sim$3\,--\,5\,MK, near the peak emission from a typical active region. Small-scale heating events would result in temporal fluctuations of the intensity of Fe~XVII emission. To generate light curves of the Fe~XVII intensity from the \magixs\ data, we sum every 4 frames of the 148 frames of \magixs\ flight data to build sufficient statistics, resulting in 144 spectroheliograms, each with an effective exposure time of 8~seconds. We then perform an inversion for each spectroheliogram and derive Level~2.0 spectrally pure maps. Figure~\ref{fig:time_evolution} shows the light curve of Fe~XVII for the XBP-1 region. We observe steady intensity of Fe~XVII emission over the entire \magixs\ flight with no sudden brightness or variations observed in the light curve. This temporal stability implies that the \textbf{XBP-1 did not encounter any sudden burst of impulsive energy release to heat the ambient plasma} during the \magixs\ observations. \\

\noindent \textbf{IV. Presence of non-Maxwellian electron distributions:} 
The non-Maxwellian $\kappa$-distribution contains Maxwellian core electrons and high-energy power-law tail. \citet{Jaro2019}  demonstrated that numerous Fe~XVII and Fe~XVIII lines within \magixs\ wavelength range are sensitive to $\kappa$ and can serve as spectral diagnostic for non-Maxwellian electrons. To study this, we first created a \magixs\ response function with $\kappa$ distributions using a $\kappa$ database \citep{Dzif2015, Dzif2021} assuming coronal abundances \citep{Feldman92}. We then performed an inversion on the \magixs\ data, solving for T and EM for different $\kappa$ values. For illustration, we considered inversions for $\kappa$ = 2 and 5, respectively. 

\begin{figure}[h!]
    \centering
    \includegraphics[width=.9\textwidth]{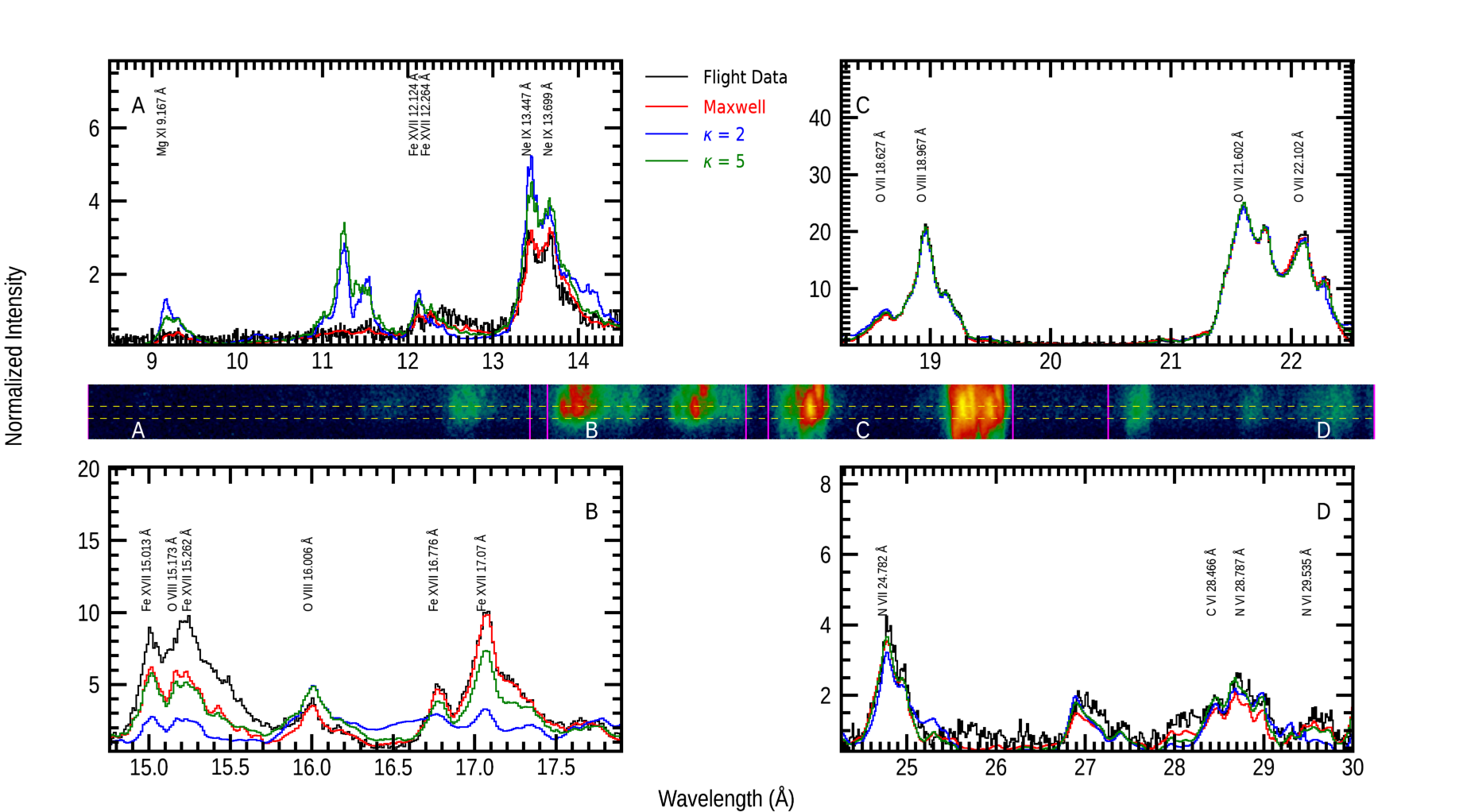}
    \caption{Comparison of flight and inverted spectra using Maxwellian and $\kappa$ = 2, and $\kappa$ = 5 electron distributions for the XBP-1 region. Top and bottom panels (A, B, C, and D) compare spectra from different wavelength ranges with the prominent emission lines labeled. The cropped flight spectroheliogram is shown in the middle with the corresponding wavelength ranges marked with vertical lines. Horizontal dashed lines indicate the rows that are summed for the XBP-1 spectra.}
    \label{fig:kappa_comp}
\end{figure}

Figure~\ref{fig:kappa_comp} shows the inverted spectroheliogram spectra for different $\kappa$ values, which indicates that the inversion with the $\kappa$ distributions does not match flight observations. Note also that an increase in $\kappa$ results in a closer match between the strong lines and the flight data, supporting the notion that increasing $\kappa$ approaches a Maxwellian electron distribution. Interestingly, we also observe that invoking a $\kappa$ distribution predicts many diagnostic emission lines that are not observed in the flight data, which could be a useful diagnostic for future flights. From these analyses we infer that the XBP-1 under study is dominated with thermal electrons with no significant non-thermal electrons, strongly suggesting that \textbf{the XBP-1 region is in thermal equilibrium}. \\

\pagebreak
\section{\label{conclusion}Conclusions}

Despite the technical challenges encountered during this first flight of the \magixs\ sounding rocket experiment that resulted in less than ideal pointing, the soft X-ray spectral images of coronal activity captured and the spectrally pure maps subsequently produced represent a revolutionary breakthrough in the field of high energy spectral imaging.  Even with the brevity of the flight and the lack of primary target observations from vignetting, \magixs\ still made a key discovery - namely, the presence of excess emission in the 15\,--\,15.7\,\AA\ wavelength range due to unmodeled Fe XVI satellite lines.  The data also strongly suggest that the observed X-ray bright point regions beyond the active region core are in thermal equilibrium and experience high-frequency (e.g., wave) heating. A re-flight of the \magixs\ instrument, with a modified configuration significantly reducing the impact of alignment tolerances, is in preparation for capturing observations of an active region core. These first results demonstrate that SXR spectral imaging can be a powerful tool in discriminating heating frequency within different coronal structures, paving the way for mapping coronal heating sources across the Sun.

\acknowledgements

We acknowledge the Marshall Grazing Incidence X-ray Spectrometer (\magixs) instrument team for making the data available through the 2014 NASA Heliophysics Technology and Instrument Development for Science (HTIDS) Low Cost Access to Space (LCAS) program, funded via grant NNM15AA15C. MSFC/NASA led the mission with partners including the Smithsonian Astrophysical Observatory, the University of Central Lancashire, and the Massachusetts Institute of Technology.  \magixs\ was launched from the White Sands Missile Range on 2021 July 30. The \magixs\ team gratefully acknowledges the \hinode\ and \iris\ (IHOP 421), \nustar, and \sdo\ missions for providing coordinated observations during the launch.

\clearpage

\bibliography{magixs_refs}{}
\bibliographystyle{aasjournal}

\end{document}